\documentclass[superscriptaddress,twocolumn,10pt,pra,aps,showpacs,longbibliography]{revtex4-1}
\usepackage{float}
\floatstyle{boxed}
\usepackage{wrapfig}
\usepackage{array}
\usepackage{graphicx}
\usepackage{amsbsy}
\usepackage{xcolor}
\usepackage[utf8x]{inputenc}
\usepackage{epstopdf}
\usepackage{amsmath,amssymb,amsfonts}
\def \am{\hat a_1}
\def \ap{\hat a_1^{\dagger}}

\newcommand{\PT}{${\cal PT}$}

\newcommand{\figref}[1]{\mbox{Fig.~\ref{#1}}}

\renewcommand{\eqref}[1]{\mbox{Eq.~(\ref{#1})}}

\newcommand{\be}{\begin{equation}}
\newcommand{\ee}{\end{equation}}
\newcommand{\bea}{\begin{eqnarray}}
\newcommand{\eea}{\end{eqnarray}}

\usepackage{url}
\usepackage[colorlinks]{hyperref}
\hypersetup{%
    plainpages=true,
    breaklinks=true,
    hypertexnames=false,
    pageanchor=true,
    colorlinks=true,
    linkcolor={blue},
    citecolor={red},
    urlcolor={blue},
    anchorcolor={black}
}

\begin{document}

\author{Ievgen I. Arkhipov}
\affiliation{RCPTM, Joint Laboratory of Optics of Palack\'y
University and Institute of Physics of CAS, Faculty of Science,
Palack\'y University, 17. listopadu 12, 771 46 Olomouc, Czech
Republic}

\author{Adam Miranowicz}
\affiliation{Faculty of Physics, Adam Mickiewicz University,
PL-61-614 Poznan, Poland} \affiliation{Theoretical Quantum Physics
Laboratory, RIKEN Cluster for Pioneering Research, Wako-shi,
Saitama 351-0198, Japan}

\author{Omar Di Stefano}
\affiliation{Theoretical Quantum Physics Laboratory, RIKEN Cluster
for Pioneering Research, Wako-shi, Saitama 351-0198, Japan}

\author{Roberto  Stassi}
\affiliation{Theoretical Quantum Physics Laboratory, RIKEN Cluster
for Pioneering Research, Wako-shi, Saitama 351-0198, Japan}

\author{Salvatore Savasta}
 \affiliation{Theoretical Quantum Physics
Laboratory, RIKEN Cluster for Pioneering Research, Wako-shi,
Saitama 351-0198, Japan}
\affiliation{Dipartimento di Scienze Matematiche e Informatiche,
Scienze Fisiche e  Scienze della Terra, Universit\`{a} di Messina,
I-98166 Messina, Italy}

\author{Franco Nori}
\affiliation{Theoretical Quantum Physics Laboratory, RIKEN Cluster
for Pioneering Research, Wako-shi, Saitama 351-0198, Japan}
\affiliation{Physics Department, The University of Michigan, Ann
Arbor, Michigan 48109-1040, USA}

\author{\c{S}ahin K. \"Ozdemir}
\affiliation{Department of Engineering Science and Mechanics, and
Materials Research Institute (MRI), The Pennsylvania State
University, University Park, Pennsylvania 16802, USA}

\title{Scully-Lamb quantum laser model for parity-time-symmetric
whispering-gallery microcavities: Gain saturation effects and
non-reciprocity}

\begin{abstract}
We use a non-Lindbladian master equation of the Scully-Lamb laser
model for the analysis of light propagation in a parity-time
symmetric photonic system {composed of coupled active and
passive whispering-gallery microresonators.} Performing the
semiclassical approximation, we obtain a set of two nonlinear
coupled differential equations describing the time evolution of
intracavity fields. These coupled equations are able to explain
the experimentally-observed light non-reciprocity [Peng {\em et
al.}, Nature Physics {\bf 10}, 394 (2014), Chang {\em et al.},
Nature Photonics {\bf 8}, 524 (2014)]. {We show that this
effect arises from the interplay between gain saturation in the
active microcavity, intercavity coupling, and losses in the cavities.}
Additionally, using this approach, we  study the effect of the
gain saturation on exceptional points, i.e., exotic degeneracies
in non-Hermitian systems. Namely, we demonstrate that the
inclusion of gain saturation  leads to a modification of the
exceptional points in the presence of intense intracavity fields.
The Scully-Lamb master equation for systems of coupled optical
structures, as proposed and applied here, constitutes a promising
tool for the study of quantum optical effects in coupled systems
with losses, gain, and gain saturation.
\end{abstract}

\date{\today}

\maketitle

\section{Introduction}

Recent years has witnessed an increasingly intense research effort
to explore a class of non-Hermitian systems described by
parity-time (\PT) symmetric Hamiltonians \cite{Bender1998}, (for
reviews see~\cite{Bender2007,Bender2003}). A system described by
the Hamiltonian $H$ is \PT-symmetric if it is invariant under the
combined action of the parity ${\cal P}$ and the time-reversal
${\cal T}$ operators (i.e., $H$ commutes with the \PT~operator:
$[H,{\cal PT}]=0$) but not necessarily with the $\cal P$ or $\cal
T$ operator alone. An important consequence of this is the
necessary, but not sufficient, condition for \PT-symmetry: The
complex potential $V(x)=V_{\rm r}(x)+i V_{\rm i}(x)$ of the
Hamiltonian should satisfy $V(x)=V^{\ast}(-x)$, where the
superscript $\ast$ denotes complex conjugation. In other words,
the real part of the potential should be an even function of $x$,
while its imaginary part should be an odd function of $x$, i.e.,
$V_{\rm r}(x)=V_{\rm r}(-x)$ and $V_{\rm i}(x)=-V_{\rm i}(-x)$. A
\PT-symmetric system exhibits two very distinct phases: an
unbroken \PT~phase (also known as the exact-\PT~regime), where the
Hamiltonian supports real eigenvalues despite being non-Hermitian,
and a broken \PT~phase where some eigenvalues form complex
conjugate pairs. The transition between these two phases takes
place spontaneously as a result of parametric variation of the
Hamiltonian. This \textit{real-to-complex spectral phase
transition} or the \textit{\PT-phase transition point} exhibits
all properties of an exceptional point (EP), which is defined as a
singularity in the parameter space of a non-Hermitian system at
which two or more eigenvalues and their associated eigenvectors
coalesce.

A decade after the ground-breaking work of Bender and Boettcher in
1998 which initiated the mathematical framework and fundamental
understanding of \PT-symmetric systems \cite{Bender1998}, it was
realized that \PT-symmetry and its breaking at an EP can be
observed in photonics by imposing the necessary condition for
\PT-symmetry on a complex optical potential, that is on the
complex refractive index, $n(x)=n_{\rm r}(x)+i n_{\rm i}(x)$,
which leads to $n_{\rm r}(x)=n_{\rm r}(-x)$ and $n_{\rm
i}(x)=-n_{\rm i}(-x)$ \cite{Ramy2007,Kostas2008}. Thus, an optical
system with the \PT-symmetric potential has a symmetric index
profile but an asymmetric gain/loss profile. Such a refractive
index profile can be obtained in two coupled optical structures,
such as waveguides or resonators: one having loss and the other
having gain compensating the loss of the other. This discovery
opened a very fertile research direction, where the interplay
between gain, loss, and the strength of the coupling between them
provides entirely new features and device functionalities
\cite{Liang2017,Ramy2018,Miri2019,Ozdemir2019}. {In these
non-Hermitian systems, with coupled loss and gain components, EPs
can be observed by controlling (or tuning) the coupling strength
to balance amplification (gain) and dissipation (loss).} Thus, EPs
can drastically alter the overall response of the system. This
leads to a plethora of nontrivial phenomena
\cite{Liang2017,Ramy2018,Miri2019}, such as enhanced light-matter
interactions \cite{Liu2016,Chen2017,Hoda2017}, unidirectional
invisibility \cite{Lin2011,Regen2012}, lasers with enhanced mode
selectivity \cite{Feng2014,Hodaei2014}, low-power nonreciprocal
light transmission \cite{Peng2014,Chang2014}, loss-induced lasing
\cite{Brands2014a,Peng2014a}, thresholdless phonon lasers
\cite{Jing2014,Lu2017} to name a few. In parallel to these efforts
in photonics, the concepts of \PT-symmetry have been put into use
in electronics \cite{Schindler2011}, optomechanics
\cite{Jing2014,Harris2016,Jing2017}, acoustics
\cite{Zhu2014,Alu2015}, plasmonics \cite{Benisty2011}, and
metamaterials \cite{Kang2013}. More recently, there is a trend in
investigating the features of \PT-symmetric quantum systems and
the effect of \PT-symmetry and its breaking on the quantum states
of light and the properties of quantum
information~\cite{Fernando2018, Kawabata2017,Ashida2017,Liu2019}.

Although \PT-symmetry and related concepts have their roots in
quantum field theories, the experimental demonstrations and the
majority of theoretical works are focused on \emph{classical}
systems, as in the example of two coupled optical microresonators:
where the energy loss in one of them  is compensated by the gain
in the other, and the system is probed with light from a laser. In
experiments, gain in such systems can be provided optically via
parametric gain or Raman gain of the {amplifying} material; while
the resonator is made from  emitters or rare-earth ions embedded
in it \cite{Peng2014,Chang2014}. The theoretical framework to
analyze such systems relies on  linearly-coupled rate equations of
classical fields, where the gain and loss correspond to a
different  sign of the imaginary part of the complex frequencies
(e.g., minus for gain and plus for loss, or vice  versa)
\emph{without} a reference to how the gain and loss are generated.
For example, the theoretical model, {developed in
Ref.~\cite{Peng2014}, was linear without any nonlinearity,
although the experimental results reported in that paper showed}
the presence of nonlinearity leading to nonreciprocal light
transmission. Thus, the theoretical framework, applied there,
failed to describe the observed non-reciprocity. On the other
hand, in Ref.~\cite{Chang2014}, closely following
Ref.~\cite{Peng2014}, a term was added phenomenologically (i.e.,
by hand) to the linear rate equations to include the effect of
gain saturation, which provides the required nonlinearity for the
nonreciprocal light transmission, \emph{without} explicitly
describing where this term comes from. { One is also able to incorporate a nonlinear term in the rate equations to
explain light non-reciprocity, but in the \PT-symmetric system of coupled
waveguides~\cite{Ramezani2010}, by resorting to a {\it semi-classical}
Maxwell-Bloch approximation, where the dynamics is governed by a
spatial variable, not time. On a different note, as the optical field is shifting
from the classical to the quantum realm, it is important that the rate
equations for the field operators are studied and the quantum-mechanical
origins of gain and loss are properly described and incorporated into the
models. Thus, a theoretical framework that addresses these concerns
is highly desirable. Moreover, since the EP depends sensitively on the
nonlinear coefficients, we can argue that having an ab-initio model is
important for the question that this paper addresses. } For this, we revisit the system,
studied in Ref.~\cite{Peng2014}, and later on in
Ref.\cite{Chang2014}, of two coupled optical structures with loss,
gain, and gain saturation by using a non-Lindbladian master
equation originally derived for the Scully-Lamb laser
model~\cite{ScullyLambBook,YamamotoBook,OrszagBook}. We apply this
master equation for the density operators of the optical fields in
optical structures. Our approach explains (at a fundamental level)
the results obtained in Refs.~\cite{Peng2014,Chang2014} and
explicitly describes {the interplay of loss, gain, and gain
saturation}.

The paper is organized as follows. Section~II contains the
Hamiltonian of the system and the Scully-Lamb laser quantum master equation
for the density operators of the optical fields. Section~III
presents the results, which include the rate equations of the
field operators and their classical limit showing explicitly the
presence of gain saturation nonlinearity and its effect on the
eigenfrequencies and exceptional points of the system, as well as
on the transmission spectra. Sections~IV and~V include a
discussion of our results, future prospects, and a summary of the
findings of this study.

\section{Laser quantum master equation for optical fields
in coupled active-passive microresonators}

\begin{figure}[tb!] 
\includegraphics[width=0.35\textwidth]{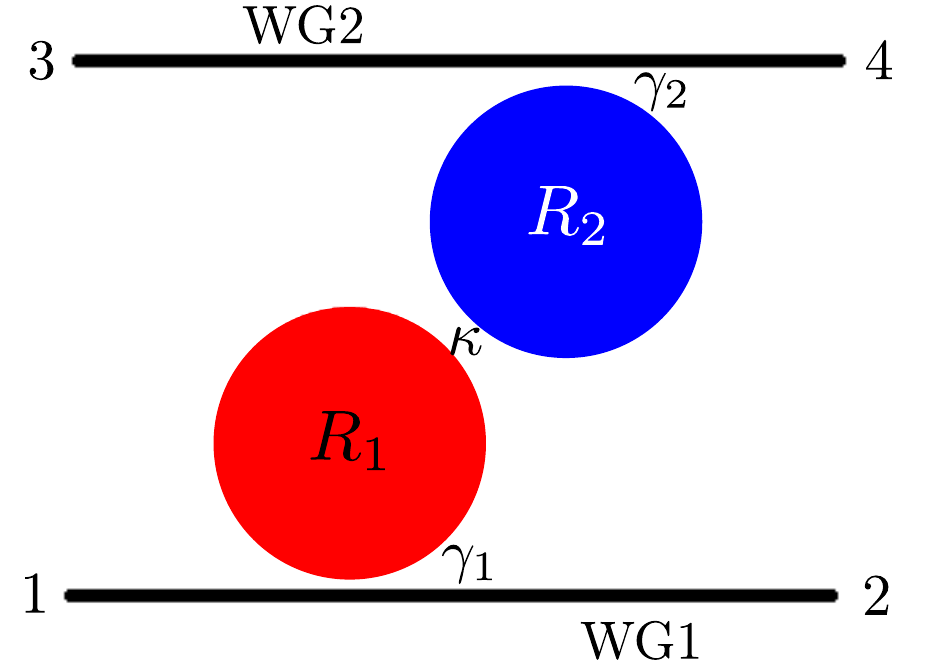}
\caption{Setup of the coupled active and passive
whispering-gallery microresonators system {studied here and in
Ref.~\cite{Peng2014}}. The active microresonator $R_1$ is coupled
to the waveguide WG1 and also coupled  to the passive resonator
$R_2$ with coupling strengths $\gamma_1$ and $\kappa$,
respectively. The coupling strength between the passive
microresonator $R_2$ and the waveguide WG2 is denoted as
$\gamma_2$. The probe signal can be input from any port labelled
from 1 to 4.}\label{fig01}
\end{figure}

Let us start from a description of the physical system that we
would like to investigate {in the Scully-Lamb laser
model~\cite{ScullyLambBook,YamamotoBook,OrszagBook}. Our interest
is focussed on the} system which was experimentally studied first
in Ref.~\cite{Peng2014} and later on in~\cite{Chang2014}. This
system consists of two coupled whispering-gallery microresonators
$R_1$ and $R_2$, where $R_1$ is an active microresonator used for
the amplification of optical fields, and $R_2$ is a passive
microcavity, which only damps the propagating fields. We denote
the $Q$-factors of the first and second microresonator as $Q_1$
and $Q_2$, respectively. The schematic diagram of the described
system is shown  in Fig.~\ref{fig01}. The coupling strength
between the two microresonators is characterized by the
real-valued parameter $\kappa$. Moreover, each microresonator is
coupled to a different waveguide denoted as WG1 and WG2 with
coupling constants $\gamma_1$ and $\gamma_2$, respectively.
{The system shown in Fig.~\ref{fig01} is $\cal PT$-symmetric
for balanced gain and loss. This is possible because the
microresonators $R_1$ and $R_2$ become interchanged under the
parity reflection $\cal P$, while loss and gain are interchanged
under the time-reversal operation $\cal T$.}

The active microresonator $R_1$ can be considered as a laser
system with a laser gain medium. Naturally, in order to describe
the dynamics of the electromagnetic field in the active laser
microresonator one would resort to the quantum laser theory and its master
equation~\cite{ScullyLambBook,YamamotoBook,OrszagBook}.

The Hamiltonian of the coupled microresonators with one driving
coherent field,  which drives the active cavity, can be
written as
\begin{equation}\label{H}  
\hat H = \sum\limits_{k=1}^2\hbar\omega_k\hat a_k^{\dagger}\hat
a_k+i\hbar\left[\kappa\hat a_1\hat a_2^{\dagger}+\epsilon\hat
a_1e^{i\omega_lt} -{\rm h.c.}\right],
\end{equation}
where {$\hat a_k$ ($\hat a_k^\dagger$) is the boson annihilation
(creation)} operator of the mode $k=1,2$, with frequency
$\omega_k$; and h.c. denotes Hermitian conjugate. {Moreover,
$\kappa$ is the coupling strength between the microresonators, and
$\epsilon$ is the coupling strength between the active cavity and
the input signal driving field, with power ${P}$ and frequency
$\omega_l$, which are related by
$\epsilon\equiv\sqrt{\gamma_1{P}/(\hbar\omega_l)}$. For the case
when the passive cavity is driven, it is enough to swap the boson
operator $\hat a_1$ with $\hat a_2$ in the last term of the
Hamiltonian in Eq.~(\ref{H}).}

By following the analysis of Yamamoto and
Imamo\v{g}lu~\cite{YamamotoBook}, {we consider our system as an
ideal laser system in the Scully-Lamb laser model, which can be
described by a non-Lindbladian master equation. This equation} for
the density operator $\hat\rho$ of the optical fields and for the
Hamiltonian $\hat H$ reads as follows (see Appendix A, for
details):
\begin{eqnarray}\label{MES}  
\frac{d}{dt}\hat\rho&=&\frac{1}{i\hbar}\left[\hat H,\hat\rho\right] +\Big[\frac{A}{2}(\ap\hat\rho\am-\am\ap\hat\rho) \nonumber \\
&&+\frac{B}{8}\Big(\hat\rho(\am\ap)^{2}+3\am\ap\hat\rho\am\ap-4\ap\hat\rho\am\ap\am\Big) \nonumber \\
&&+\sum\limits_{i=1}^2\frac{\Gamma_i}{2}(\hat a_i\hat\rho\hat
a_i^{\dagger}-\hat a_i^{\dagger}\hat a_i\hat\rho) + \rm {h.
c.}\Big],
\end{eqnarray}
where the gain $A$ and  gain saturation $B$ coefficients for the field
in the active cavity are expressed as
\begin{equation}\label{AB}  
A=\frac{2g^2r}{Y^2}, \quad \text{and} \quad B=\frac{4g^2}{Y^2}A,
\end{equation}
respectively.  The parameter $g$ stands for the coupling strength
between the atoms of the gain medium and the optical field in the
active cavity, $Y$ is a decay rate of the atoms, and $r$ accounts
for the pump rate of the gain medium.  In \eqref{MES}, the
decaying rates for both cavities are denoted by ($i=1,2$)
\begin{eqnarray}  
\Gamma_i=C_i+\gamma_i, \quad {\rm where} \quad
C_i=\frac{\omega_i}{Q_i} \label{N13}
\end{eqnarray}
is the intrinsic loss of the $i$th cavity, and $\gamma_i$ stands
for the loss due to the coupling of the $i$th cavity to the $i$th
waveguide.

{We note that the derivation of the master equation in
Eq.~(\ref{MES}), as carried out in Appendix~A, is based on two
main assumptions: (i) the adiabatic elimination of the population
in the gain medium of the active resonator, and (ii) the
weak-gain-saturation regime, i.e., when the laser in the active
cavity operates not far from the lasing threshold. The
non-Linbladian form of the master equation in Eq. (\ref{MES}),
which is obtained within the fourth-order field
approximation~\cite{YamamotoBook,OrszagBook}, is due to quantum
 jump-operator terms, which account for gain saturation. }

We also note that the master
equation, given in \eqref{MES}, can be {recast} to the
Lindbladian form as~\cite{Gea1998}:
\begin{equation}\label{Lindblad}  
\frac{d}{dt}\hat\rho=\frac{1}{i\hbar}\left[\hat H,\hat\rho\right]
-\frac{1}{2}\sum\limits_{i=1}^4\left(\hat L_i^{\dagger}\hat
L_i\hat\rho+\hat\rho\hat L_i^{\dagger}\hat L_i-2\hat
L_i\hat\rho\hat L_i^{\dagger}\right),
\end{equation}
where the Lindblad operators $\hat L_i$ (for $i=1,\dots,4$) are
defined as:
\begin{eqnarray}  
&\hat L_1 = \sqrt{A}\ap\left(1-\frac{B}{2A}\am\ap\right), \quad \hat L_2 = \frac{1}{2}\sqrt{3B}\am\ap, &\nonumber  \\
&\hat L_3 = \sqrt{\Gamma_1}\am, \quad \hat L_4 =
\sqrt{\Gamma_2}\hat a_2.&
\end{eqnarray}
The Lindblad form in \eqref{Lindblad} is equivalent to the master
equation~(\ref{MES}) if the terms of second order in
$B\am\ap/(2A)$ are neglected in Eq.~(\ref{Lindblad}), which holds true for the weak gain saturation regime.

\section{Results}

\subsection{Rate equations}

 The master equation for the field operators, given in
\eqref{MES}, and the Hamiltonian in \eqref{H},  yield the rate
equations for the averaged boson operators $\am$ and $\hat a_2$.
Namely, by using the formula
\begin{equation} \label{REF} 
\frac{d}{dt}\langle \hat a_j\rangle = {\rm Tr}\left[\hat
a_j\frac{d}{dt}\hat\rho\right], \quad j=1,2,
\end{equation}
and utilizing the cyclic property of the trace operation, after
substituting \eqref{MES} in \eqref{REF}, we obtain
\begin{eqnarray}\label{RE}  
\frac{d}{dt}\langle \hat a_1\rangle &=&-iw_1\langle \hat a_1\rangle+\frac{G_1}{2}\langle \hat a_1\rangle-\kappa\langle \hat a_2\rangle -\frac{B}{2}\langle \ap\am\am\rangle, \nonumber \\
&&-\epsilon \exp\left({-i\omega_lt}\right), \nonumber \\
\frac{d}{dt}\langle\hat a_2\rangle &=&-iw_2\langle \hat
a_2\rangle-\frac{\Gamma_2}{2}\langle \hat a_2\rangle+\kappa\langle
\hat a_1\rangle,
\end{eqnarray}
where $G_1=A-\Gamma_1-\frac{7}{4}B$.

As one can see, the rate equations, given in \eqref{RE} for the
averaged quantum amplitudes, are nonlinear due to the presence of
the gain saturation in the active cavity, i.e., the term
$(B/2)\langle \ap\am\am\rangle$ represents the nonlinearity.

\subsection{Classical Limit}

In the classical limit, i.e., in the case of large intensities of
the fields, the quantum field operators can be represented by
$c$-number amplitudes as  $\hat a_i\rightarrow \langle \hat
a_i\rangle\equiv a_i$. Then the rate equations in \eqref{RE} can
be rewritten in the classical limit as
\begin{eqnarray}\label{REC1} 
\frac{d}{dt}a_1 &=&-iw_1a_1+\frac{G_1}{2}a_1-\kappa a_2 -\frac{B}{2}|a_1|^2a_1-\epsilon  \exp\left({-i\omega_lt}\right), \nonumber \\
\frac{d}{dt}a_2 &=&-iw_2a_2-\frac{\Gamma_2}{2}a_2+\kappa a_1,
\end{eqnarray}
where the term $\frac{B}{2}|a_1|^2a_1$ represents the nonlinearity
due to gain saturation. By substituting  $a_k=A_k(t)\exp({-i\omega
t})$ into \eqref{REC1}, one arrives at
\begin{eqnarray}\label{REC}  
\frac{d}{dt}A_1 &=&i\Delta A_1+\frac{G_1}{2}A_1-\kappa A_2 -\frac{B}{2}|A_1|^2A_1-\epsilon, \nonumber \\
\frac{d}{dt}A_2 &=&i\Delta A_2-\frac{\Gamma_2}{2}A_2+\kappa A_1,
\end{eqnarray}
where $\Delta=\omega-\omega_c$ is the frequency-detuning
parameter, and we assumed that the frequency of the driving
coherent field is $\omega_{l}=\omega$, and the frequencies of the
cavities are the same, $\omega_1=\omega_2=\omega_c$. {We note
that when the input driving field enters the passive cavity from
port~4 in Fig.~\ref{fig01}, then the term $(-\epsilon)$ would
appear in  the second equation (instead of the first) in
Eq.~(\ref{REC}). Importantly, in the semiclassical approximation,
the coupling constant $\epsilon$ can be expressed via the complex
amplitude $A_{\rm in}$ of the input driving field as
$\epsilon\equiv\sqrt{\gamma_j}A_{\rm in}$, for $j=1,2$.
Nevertheless, for convenience, we will keep the current notation
for the input driving field via $\epsilon$, because it shares the
same dimensionality with the other parameters describing the system.
Moreover, by keeping this notation it will be more transparent and
easier for us to compare the semiclassical and fully quantum
approaches in related future research. In what follows, we will
always recall the connection between the driving coupling constant
$\epsilon$ and the input driving field amplitude $A_{\rm in}$ to
avoid any confusion.}

It is clearly seen from \eqref{REC} that the rate equations
simplify to a linear form if the  gain saturation in
the active laser cavity are neglected.

{We note that in the semiclassical Maxwell-Bloch picture,
which is widely used for describing the characteristic $\cal
PT$-symmetric properties of classical optical fields, one can
introduce gain saturation via a modified gain coefficient $A$ for
the field $A_1$ in the active cavity as $A\rightarrow
A/(1+|A_1|^2/|A_{\rm s}|^2)$ in Eq.~(\ref{REC}) for $B=0$, where
$|A_{\rm s}|^2$ is the gain saturation threshold~\cite{Chang2014}.
In this case, by decomposing this modified gain coefficient $A$ in
a Taylor series (up to the first order in $|A_1|^2/|A_{\rm s}|^2\ll
1$, which is justified in the weakly saturated regime), one
obtains a one-to-one correspondence between the semiclassical
Maxwell-Bloch picture and the semiclassical Scully-Lamb laser
model for the optical fields in Eq.~(\ref{REC}) by setting $B =
A/|A_{\rm s}|^2$. Nevertheless, compared to the Maxwell-Bloch
picture, where the gain saturation term is introduced
phenomenologically, the semiclassical Scully-Lamb laser theory
explains gain saturation not only qualitatively, but also
quantitatively, via the laser system parameters given in
Eq.~(\ref{AB}).}

{We also note that a set of nonlinear equations, similar to
Eq.~(\ref{REC}), can be obtained within the semiclassical
Maxwell-Bloch approximation for $\cal PT$-symmetric coupled
waveguides~\cite{Ramezani2010}. Nonetheless,   there are two main
differences with respect to Ref.~\cite{Ramezani2010}. The first is
that we consider the case of coupled resonators instead of coupled
waveguides. In our case, feedback effects and two different kinds
of losses (i.e., input-output and intrinsic losses) are
considered. Second, even if we apply the semiclassical
approximation, our theory is based on a fully quantum approach,
which can be useful to investigate quantum effects and quantum
noise. }

\subsection{Eigenfrequencies in the steady-state and exceptional point}

In the steady state, by considering $\gamma_1=\gamma_2=\gamma$ for
symmetry reasons, we find from \eqref{REC} that:
\bea\label{SREC}  
\left(i\Delta +\frac{G'_1}{2}\right)A_1-\kappa A_2&=&\epsilon,\nonumber \\
\left(i\Delta-\frac{\Gamma_2}{2} \right)A_2 + \kappa A_1 &=&0\,,
\eea where $G'_1=G_1-B|A_1|^2$. {Because the system of
equations in Eq.~(\ref{SREC}) is written for the complex fields
$A_j$, we can incorporate the real steady-state intensity
$|A_1|^2$ into the prefactor $G_1'$. One can obtain the
eigenvalues of the system, given in Eq.~(\ref{SREC}), by  setting
the driving field to zero ($\epsilon=0$). Defining the vector
$\alpha=(A_1,A_2)^T$, we can rewrite Eq.~(\ref{SREC}) as
follows
\begin{equation}  
i\frac{{\rm d}\alpha}{{\rm d}t}=M\alpha,
\end{equation}
with the evolution matrix
\begin{equation}\label{M}  
M=\begin{pmatrix}
-\Delta +i\frac{G'_1}{2} & -i\kappa \\
i\kappa & -\Delta-i\frac{\Gamma_2}{2}
\end{pmatrix}.
\end{equation}
For the case when gain and losses are balanced, i.e., when
 $G_1'=\Gamma_2$ holds, the evolution matrix $M$, given in
Eq.~(\ref{M}), becomes $\cal PT$-symmetric, i.e., $[M,{\cal
PT}]=0$, with the parity operator $\cal P=\begin{pmatrix}0 & 1\\1
&0\end{pmatrix}$ and $\cal T$ performing complex conjugation.}

{The characteristic equation of the matrix $M$ is}
\be  
\left(\Delta-i
\frac{G'_1}{2}\right)\left(\Delta+i\frac{\Gamma_2}{2}\right)-\kappa^2=0,
\ee from which we find {the formal solution for the eigenfrequencies, as follows}
\be  
\Delta_\pm=i \frac{G'_1-\Gamma_2}{4}\pm\sqrt{4\kappa^2
-\frac{1}{4}\left(G'_1+\Gamma_2\right)^2}\,. \ee By recalling that
$\Delta_\pm =\omega_\pm-\omega_c$, we obtain
\begin{eqnarray}\label{wpm}  
\omega_\pm&=&\omega_c+\frac{i}{4}\Big(A-C_1-C_2-BI_1-2\gamma\Big) \nonumber \\
&&\pm \frac{1}{2}\sqrt{4\kappa^2
-\frac{1}{4}\left(A-C_1+C_2-BI_1\right)^2},
\end{eqnarray}
{where $I_1=|A_1|^2$ is the dimensionless intensity of the field in
the active cavity in the steady-state, and which is by itslef a function of the frequency $\omega$ (for the solutions of the steady-state  intensity $I_1$, see Section III~E).} {The solutions in
Eq.~(\ref{wpm}) give the energy eigenspectra of the system, i.e.,
the eigenvalues of the system Hamiltonian, which determine the
evolution matrix $M$.}

\begin{figure}[tb!]  
\includegraphics[width=0.35\textwidth]{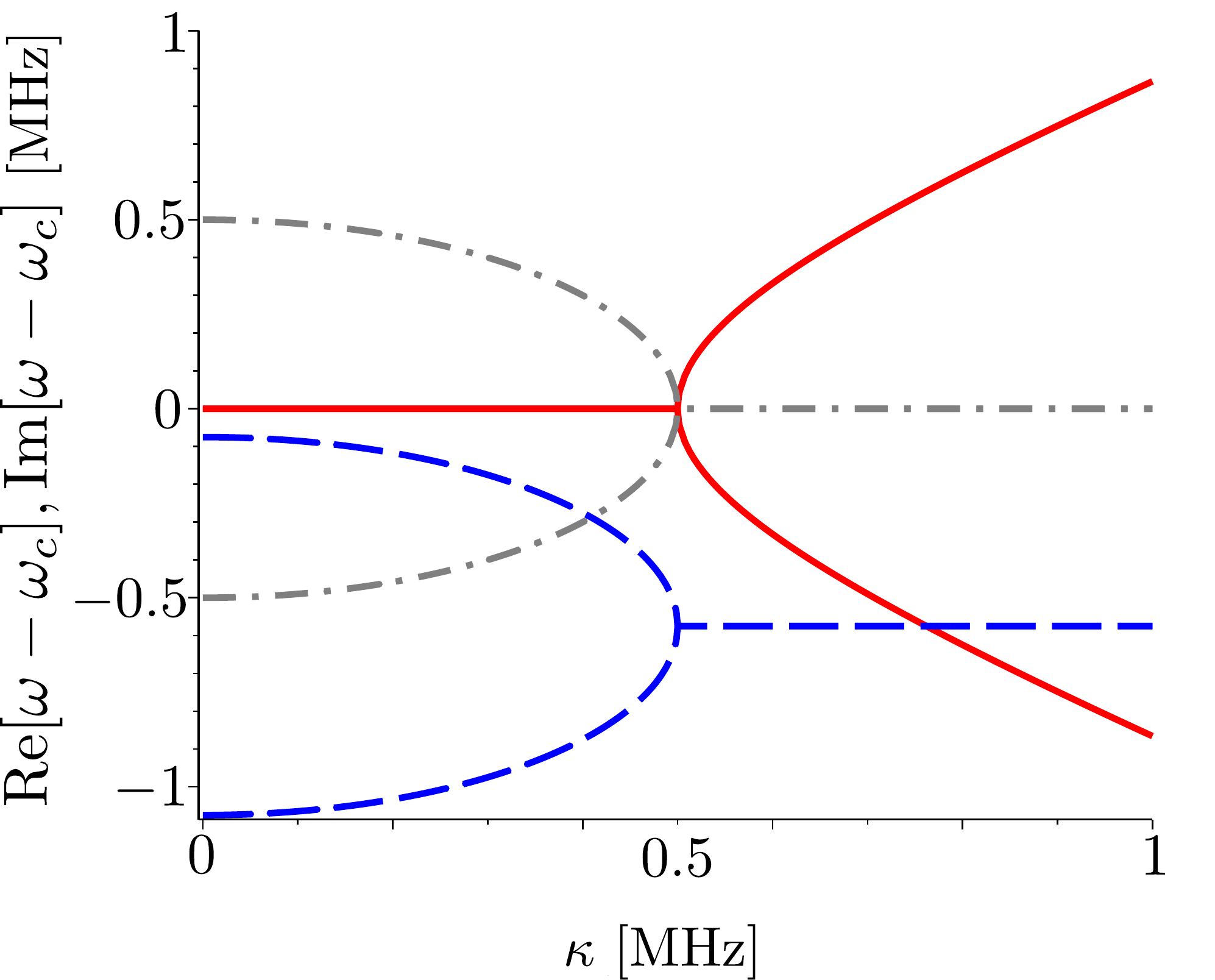}
\caption{Real part, ${\rm Re}[\omega]$, (red solid curve) and
imaginary part, ${\rm Im}[\omega]$, (blue dashed curve) of the
eigenfrequencies of the supermodes, as a function of coupling
$\kappa$ at the \PT-symmetry condition $A-C_1-C_2-BI_1=0$. The
exceptional point does not depend on gain saturation. We have
chosen $C_2=1$~MHz, $\gamma=1.15$~MHz.  In addition,  we show
${\rm Re}[\omega]$ (red solid curve, {which overlaps with the
other red curve}) and ${\rm Im}[\omega]$ (gray
dash-dotted curve) of the eigenfrequencies of the supermodes when
the input-output losses are not considered. Here $\omega_c$ stands
for the cavities resonance. \label{fig02}}
\end{figure}

We note again that, in reality, the loss rates $\gamma_i$, arising
from the coupling with the input-output channels,  are not true
losses, because these describe the energy transfer from the system
to the output (or from the input to the system). Hence, the
concept of the {\it effective} \PT-symmetry in our  system with
balanced gain and loss, can be expressed as
\be\label{pt tot}  
A-C_1-C_2-BI_1=0\, . \ee This is valid because the system is
\PT-symmetric regardless of how it is probed (i.e., waveguides in
the coupled resonator systems are used only to probe the system).
When the condition, given in Eq.~(\ref{pt tot}), is satisfied we
find from \eqref{wpm} that:
\be\label{wpmPT}  
\omega^{\rm PT}_\pm=\omega_c-i \frac{\gamma}{2}\pm
\frac{1}{2}\sqrt{4\kappa^2 -C_2^2}\,. \ee

Note that, when changing the input signal
$\epsilon$ (and hence the resulting steady-state intensity $I_1$), one
has to adjust the losses correspondingly in order to satisfy the
\PT-symmetry condition.

{The analysis of the frequency spectrum, given  in Eqs.~(\ref{wpm})
and (\ref{wpmPT}), provides two different regimes,  {or so-called \emph{unbroken} and \emph{broken} $\cal PT$-symmetry phases,} depending on the
sign of the expression under the square-root sign. In the unbroken $\cal PT$-symmetry phase,  the
expression under the root is positive (that is,
$\kappa>\frac{C_2}{2}$), and there are always two supermodes with
non-degenerate real frequencies $\omega_{\pm}$ that
propagate in the system.   Note that this is true for the system
itself, where the coupling loss $\gamma$ (which is not inherent to
the system of coupled resonators) is zero (i.e., $\gamma=0$). In the broken $\cal PT$-symmetry phase,
that expression is negative (that is,  $\kappa<C_2/2$), and the real
spectrum becomes degenerate, indicating that the system displays
two modes with the same resonance frequency but with different
decay rates.} The transition between these two regimes takes place
at an EP given by
 \be \label{ptcond}  
\kappa_{\rm EP}=\frac{C_2}{2}\, .
 \ee

In \figref{fig02} we show the real and imaginary parts of the
eigenfrequencies of the supermodes. As expected by the inspection
of \eqref{ptcond}, the {EP at $\kappa=\kappa_{\rm EP}$} does
not depend on the field intensity. For $\kappa>\kappa_{\rm EP}$,
we observe that the imaginary part is different from zero. This is
due to the contributions of the loss rates $\gamma_i$ arising from
the coupling with the input-output channels. The gray dash-dotted
curve in \figref{fig02} describes the imaginary parts of the
complex eigenfrequencies, when the input-output coupling losses
are neglected (i.e., $\gamma=0$).

\begin{figure}[tb!]  
\includegraphics[width=0.5\textwidth]{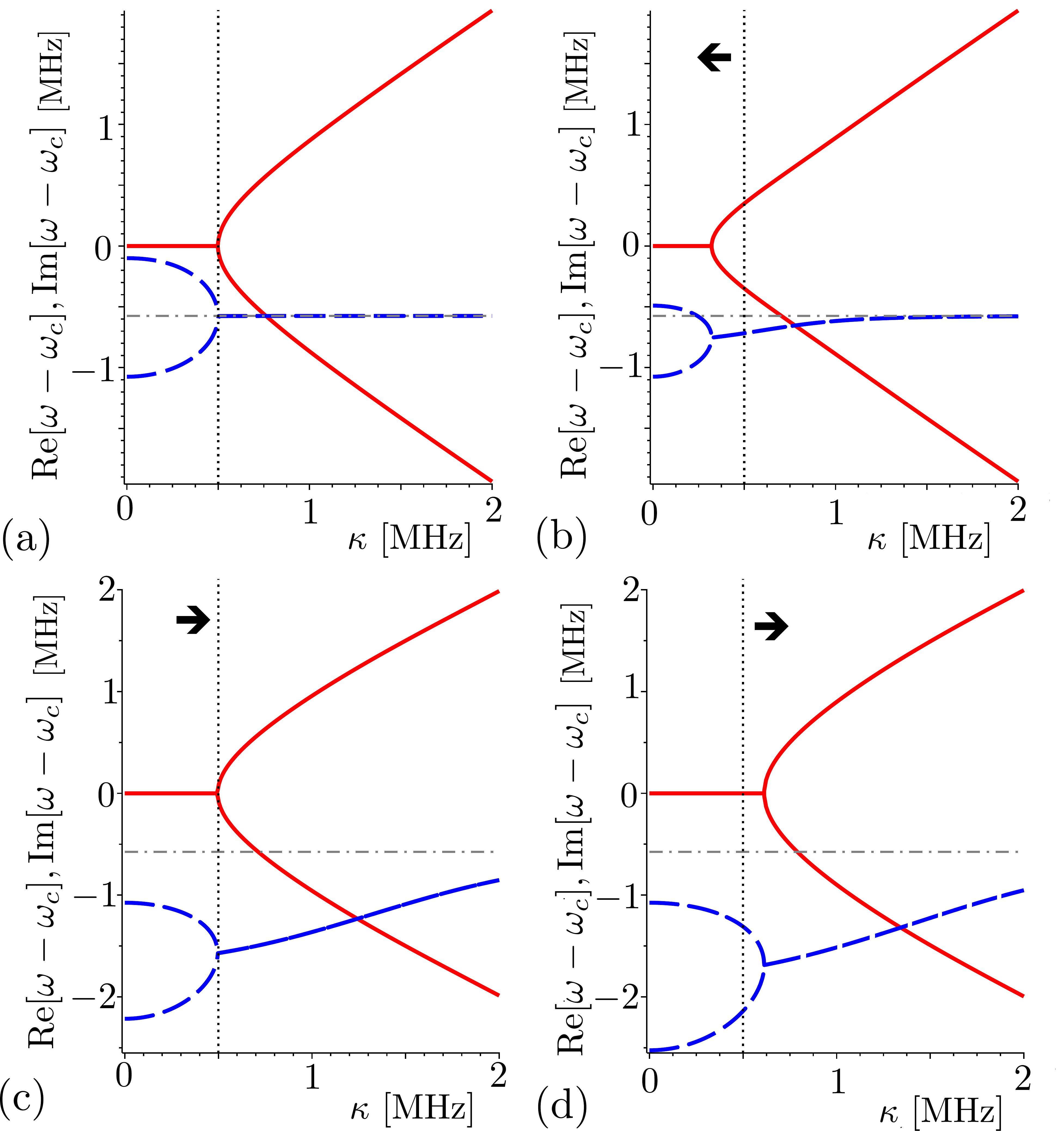}
\caption{Real part, ${\rm Re}[\omega]$, (red solid curve) and
imaginary part, ${\rm Im}[\omega]$, (blue dashed curve) of the
eigenfrequencies of the supermodes as a function of the coupling
$\kappa$ under the \PT-symmetry condition $A-C_1-C_2=0$ (i.e.,
excluding the gain saturation term $BI_1$) for different values of
the coupling coefficient $\epsilon$  of the driving field to a
microresonator (see Fig.~\ref{fig01}). It is seen that exceptional
point depends on gain saturation. (a) The linear regime with
$B=0.05$~Hz and $\epsilon=1$ MHz; and (b, c, d) the nonlinear
regime according to \eqref{NPT} with $B=0.05$~Hz with (b)
$\epsilon=2$~GHz, (c) $\epsilon=20.5$~GHz, and (d)
$\epsilon=25$~GHz. The observed inclination of the imaginary part
of the eigenfrequencies (blue dashed curve) towards negative
values near the EP indicates an additional loss due to gain
saturation caused by stronger driving fields. We assumed a passive
cavity loss of $C_2=A-C_1=1$~MHz, an active cavity gain of $A=301$
MHz, and a waveguide-microresonator coupling strength of
$\gamma=1.15$~MHz. The vertical black dashed line denotes the EP
$\kappa_{\rm EP}$ for the linear system, when the gain saturation
term $BI_1=0$. The horizontal gray dash-dotted line denotes the
converging value ($\gamma/2$) for the imaginary part of the
eigenfrequencies (blue dashed curve) for large $\kappa$. The
{condition $BI_1/A\ll 1$ for a weak saturation} is always
satisfied for all the cases corresponding to panels (a)--(d) (see
also Appendix~B).} \label{fig03}
\end{figure}

We now consider the case, where the  \PT-symmetry condition is
achieved at low input rates, so the  gain saturation effects
are negligible ($BI_1 \simeq 0$). In this case, as for the
\PT-symmetry condition, we can use   \eqref{pt tot} calculated for
$I_1=0$:
 \be\label{pt_tot2}  
A-C_1-C_2=0\, .
 \ee
If the input drive is increased without adjusting the other
parameters ($\kappa$ and $C_i$), the \PT-symmetry condition is
{not} satisfied any more. We will investigate the
effect of gain saturation on the spectral properties of the
considered \PT-symmetric system, which is usually studied without
 the inclusion of gain saturation. {Thus, hereafter, we
assume that the \PT-symmetric condition is given in
\eqref{pt_tot2}.} The condition for the exceptional point becomes
[see \eqref{wpm}]:
 \be\label{ptapprox}  
\kappa_{\rm EP}
=\frac{1}{4}\left|A-C_1+C_2-BI_1\right|=\frac{1}{4}\left|2C_2-BI_1\right|\,.
 \ee
From inspection of \eqref{ptapprox}, we observe that, in this
case, the EP changes when the steady-state field intensity
increases in the active cavity (see \figref{fig03}). In
particular, as an example, in \figref{fig03} we plot the real and
imaginary parts of the eigenfrequencies of the supermodes for the
linear [Fig.~\ref{fig03}(a)] and nonlinear
[Figs.~\ref{fig03}(b)--(d)] regimes for different values of the
driving-field coupling $\epsilon$ when driving the system at a
resonant frequency $\omega_c$.

\begin{figure}[tb!]  
\includegraphics[width=0.488\textwidth]{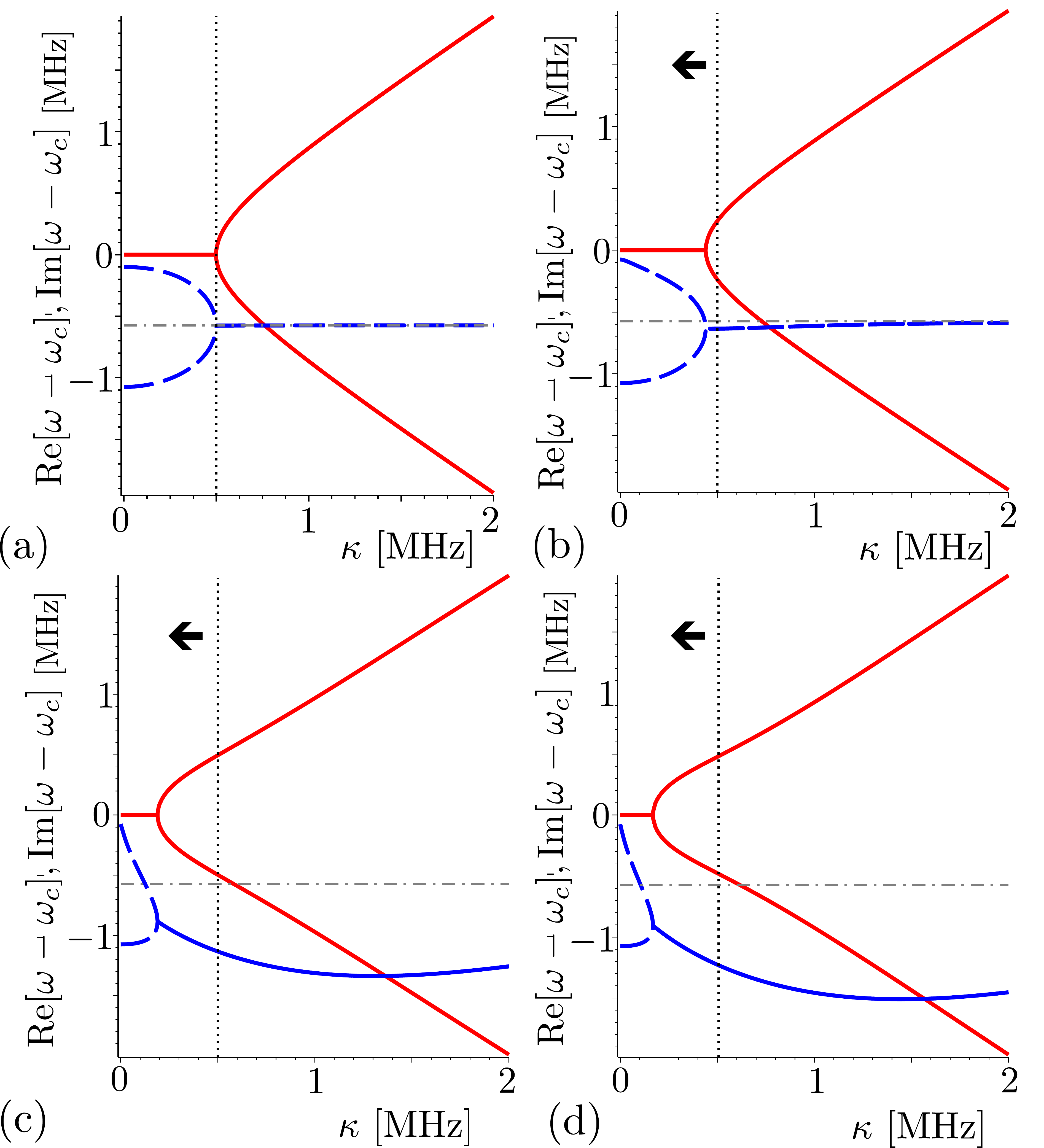}
\caption{Real part, ${\rm Re}[\omega]$, (red solid curve) and
imaginary part, ${\rm Im}[\omega]$, (blue dashed curve) of the
eigenfrequencies of the supermodes as a function of the coupling
$\kappa$ under the  \PT-symmetry condition $A-C_1-C_2=0$ (i.e.,
excluding the gain saturation term $BI_1$) for different values of
the coupling coefficient $\epsilon$  when the driving field drives
the passive cavity (see Fig.~\ref{fig01}). The parameters used here are the
same as in Fig.~\ref{fig03}. }\label{fig04}
\end{figure}
\begin{figure}[tb!]  
\includegraphics[width=0.488\textwidth]{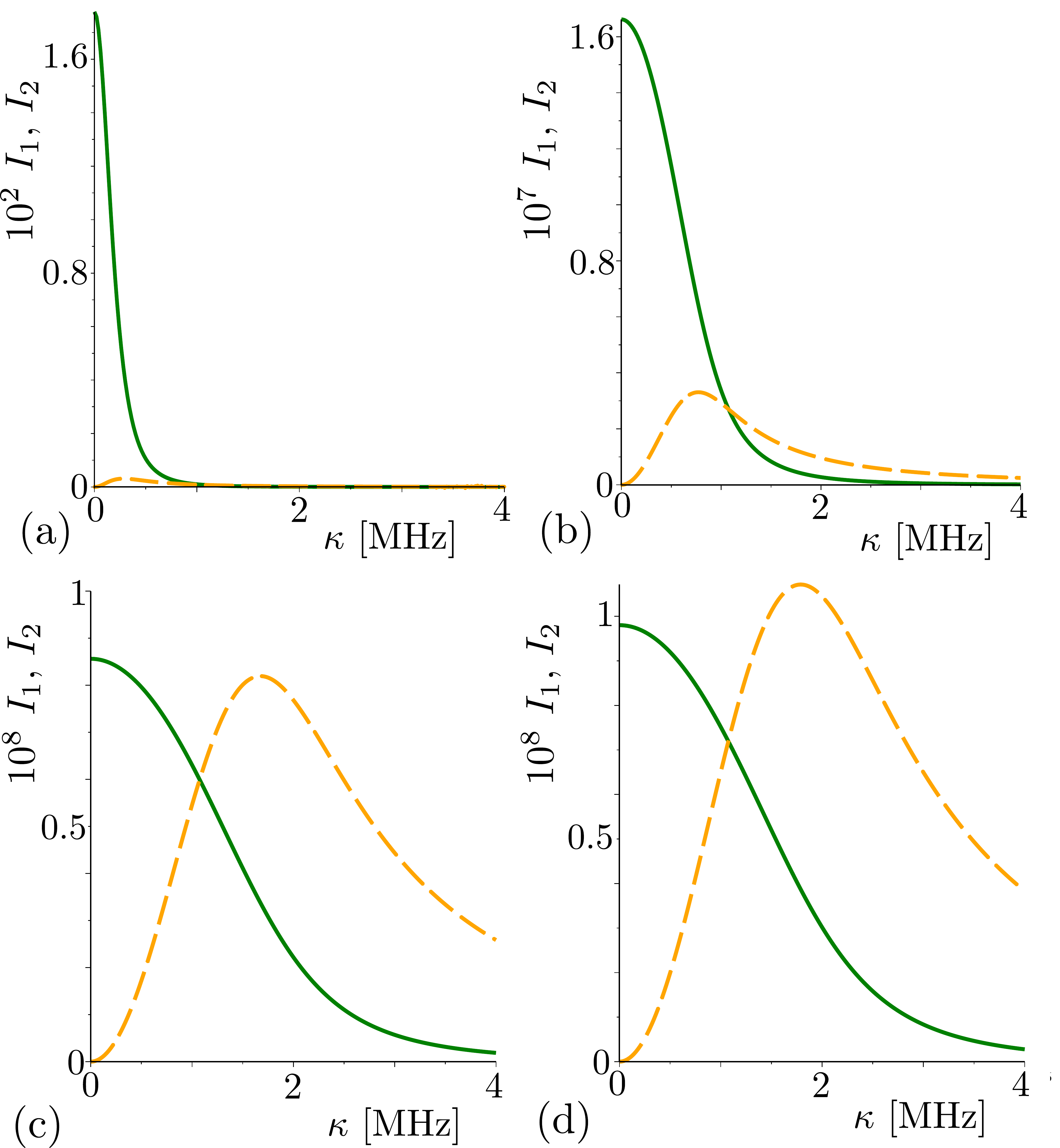}
\caption{Dimensionless steady-state intensities $I_1$ (green solid
curve) and  $I_2$ (yellow dashed curve) in the active and passive
cavities, respectively, as a function of the intercavity coupling
$\kappa$ for the driving field propagating in the direction
$1\rightarrow 4$ (see Fig.~\ref{fig01}).  The panels (a) -- (d) are obtained
for different values of the driving field coupling $\epsilon$ and
correspond to the panels in Fig.~\ref{fig03}.}\label{fig05}
\end{figure}
\begin{figure}[tb!]  
\includegraphics[width=0.488\textwidth]{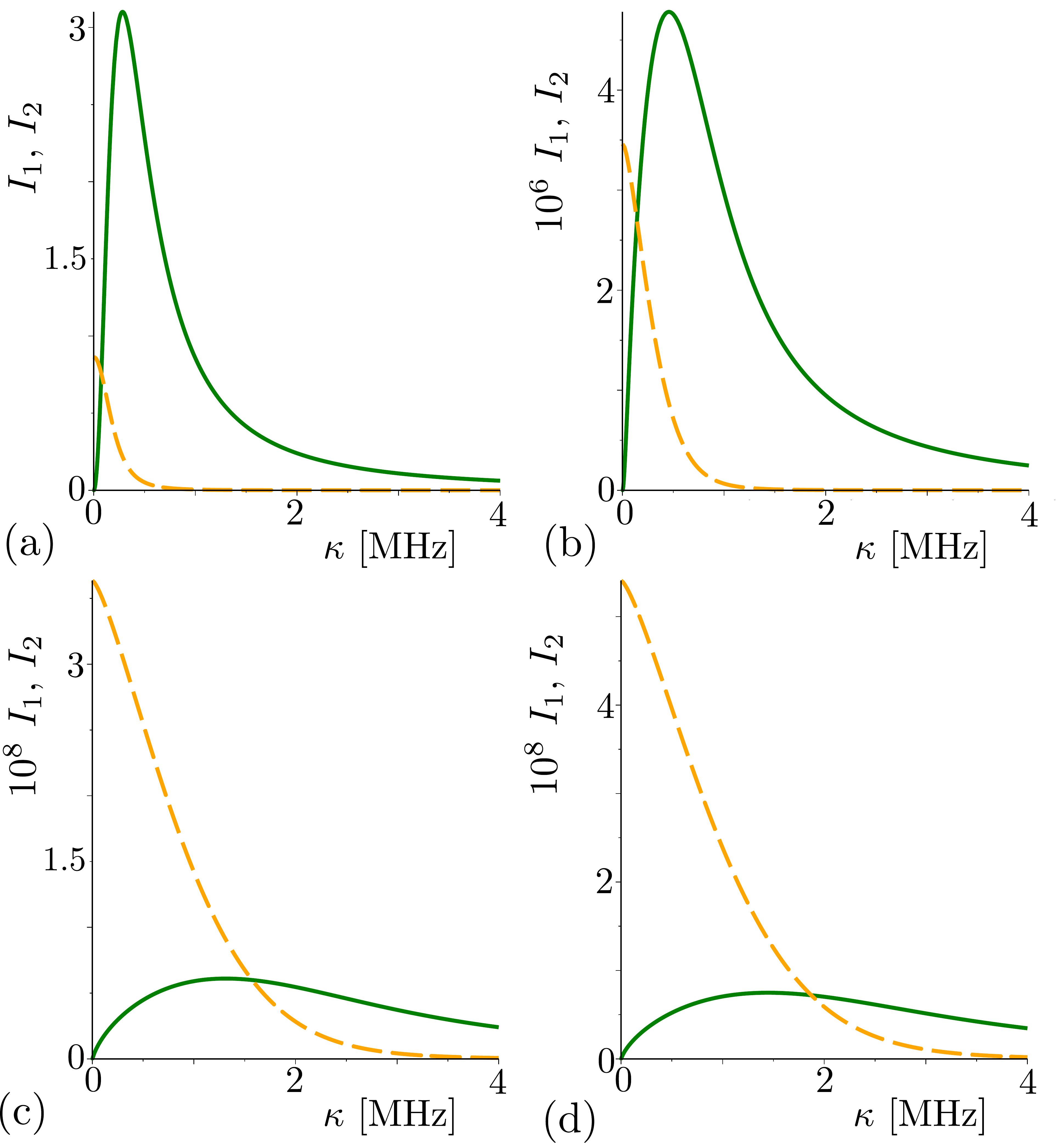}
\caption{Dimensionless steady-state intensities $I_1$ (green solid
curve) and  $I_2$ (yellow dashed curve) in the active and passive
cavities, respectively, as a function of the intercavity coupling
$\kappa$ for the driving field propagating in the direction
$4\rightarrow 1$ (see Fig.~\ref{fig01}).  The panels (a) -- (d) are obtained
for different values of the driving-field coupling $\epsilon$ and
correspond to the panels in Fig.~\ref{fig04}.}\label{fig06}
\end{figure}
In the linear regime, the gain saturation term $BI_1$ is either
exactly zero or negligible compared to the system parameters $A$
and $C_i$; thus, the behavior of the EP is identical to that
presented in Fig.~\ref{fig02} [see also Figs.~\ref{fig03}(a) and
\ref{fig04}(a)]. The same conclusion applies when the driving
field is far away from the resonance $\omega_c$, as in that case,
the steady-state intensity also tends to zero.

An interesting situation arises when the gain saturation term
$BI_1$ becomes comparable with the passive cavity loss $C_2$.
However, at the same time, it is much less than the gain
coefficient $A$ in the active cavity, i.e., $BI_1\ll A$, so the
weakly saturated regime  still holds and the validity of the
master equation in \eqref{MES} remains. This also implies that
$C_1\gg C_2$. In this case, the system starts exhibiting some
nonlinear features in its eigenspectrum. In short, the described
condition can be written as
\begin{equation}\label{NPT} 
BI_1\approx C_2\ll A\approx C_1.
\end{equation}
In what follows, we always call the system to be in the
\emph{nonlinear} \PT-symmetric regime, whenever \emph{both}
conditions, given in Eqs.~(\ref{pt_tot2}) and (\ref{NPT}), are
satisfied. In that case, one can observe that the critical value
of $\kappa_{\rm EP}$ significantly changes depending on the gain
saturation term $BI_1$. Moreover, the steady-state intensity $I_1$
in the active cavity by itself becomes dependent on the direction
of the propagation of the driving field, i.e., on whether the
driving field is coupled to the active (from port~1 to port 4) or
the passive cavity (from port~4 to port 1) [see Fig.~1].
Consequently, the gain saturation term $BI_1$ also depends on the
driving-field direction.

{When the input driving field is coupled to the active cavity,
it can experience a significant gain saturation for large input
intensities, and, as a result, its losses increase. This
especially happens when one decreases the intercavity coupling
$\kappa$.  In the latter case, the strong signal field becomes
localized in the active resonator, before being transferred to the
lossy passive cavity. This leads to the gain decrease of the
intense driving field due to gain saturation, and, therefore, to
the observed signal-field losses (see blue dashed curve in
Fig.~\ref{fig03}). Moreover, when the losses induced by the gain
saturation become comparable to the losses in the passive cavity
(i.e., when $BI_1\approx C_2$), then the critical value of
$\kappa_{\rm EP}$ first decreases and then increases when
increasing the intensity of the input field (see
Fig.~\ref{fig03}). As Eq.~(\ref{ptapprox}) implies, this shift of
$\kappa_{\rm EP}$ can be explained by the interplay between losses
in the passive cavity and losses induced by gain saturation in the
active resonator. Namely, by increasing the gain saturation term
$BI_1$ (i.e., by increasing the input signal) in the proximity of
$C_2$ in Eq.~(\ref{ptapprox}), the critical value $\kappa_{\rm
EP}$ first decreases, when $BI_1\le C_2$, and then increases, when
$BI_1\ge C_2$.}

\subsection{Non-reciprocity of light propagation}

{Another important result of our work is the theoretical
microscopic \emph{prediction of non-reciprocity of the propagating
light}. {Specifically, in broken $\cal PT$-symmetry phase, i.e., when $\kappa<\kappa_{\rm EP}$,  
the signal field can experience smaller losses for smaller values of
$\kappa$, when the driving field propagates in the
direction $4\rightarrow1$, in comparison to the case when it propagates 
in the opposite direction
$1\rightarrow4$ (see Fig.~\ref{fig04}).} This stems from the fact,
that when the strength of the input signal field is increasing, by
decreasing the intercavity coupling $\kappa$, the driving field
experiences large losses in the passive cavity before passing into
the active resonator. Now, because the initially strong input
signal field is strongly damped by the passive resonator, it
enters the active cavity having an intensity which is not
sufficient to induce gain saturation. As such, the propagating
field in the active cavity can even undergo notable amplification
before being detected at port 1. As a result, the critical values
of $\kappa_{\rm EP}$ only decrease when increasing the intensity
of the input signal field, since the propagating field cannot
reach high intensity in the active cavity to make the term $BI_1$
larger than $C_2$ in Eq.~(\ref{ptapprox}) (see also
Fig.~\ref{fig04}).}

\begin{figure}[tb!]  
\includegraphics[width=0.43\textwidth]{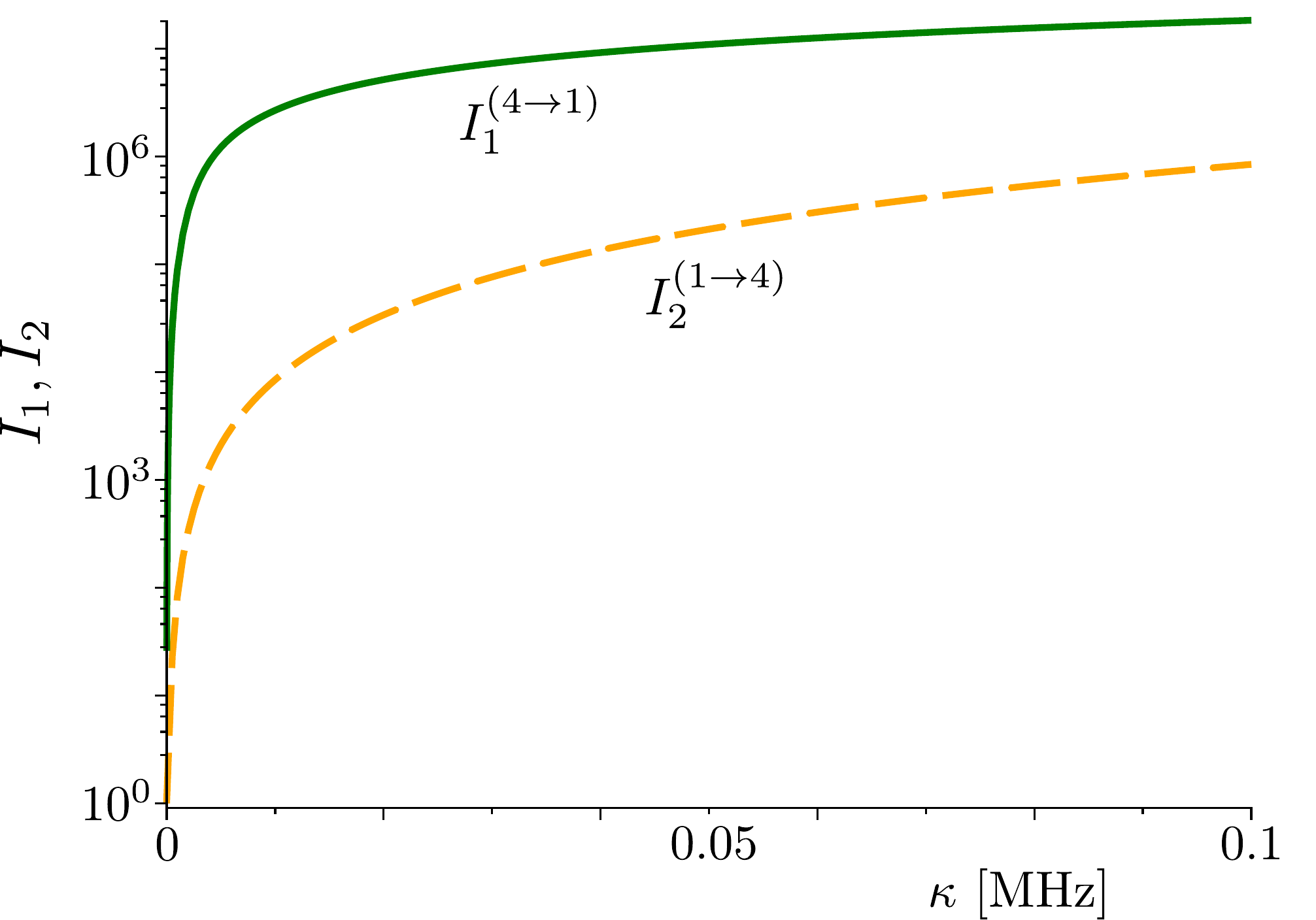}
\caption{Dimensionless steady-state intensity $I_1^{(4\rightarrow
1)}$ (green solid curve) in the active cavity for the driving
field propagating  in the direction $4\rightarrow 1$ from
Fig.~\ref{fig06}(d), and the steady-state intensity
$I_2^{(1\rightarrow 4)}$ (yellow dashed curve) in the passive
cavity for the driving field propagating in the direction
$1\rightarrow 4$ from Fig.~\ref{fig05}(d) for lower values of
$\kappa$. For the broken-\PT-symmetric phase in the nonlinear
regime for small values of $\kappa<\kappa_{\rm EP}$, the
steady-state intensity  $I_1^{(4\rightarrow 1)}$  is two-three orders of
magnitude larger than  $I_2^{(1\rightarrow 4)}$, implies
nonreciprocal light propagation.}\label{fig07}
\end{figure}

\begin{figure}[tb!]  
\includegraphics[width=0.4\textwidth]{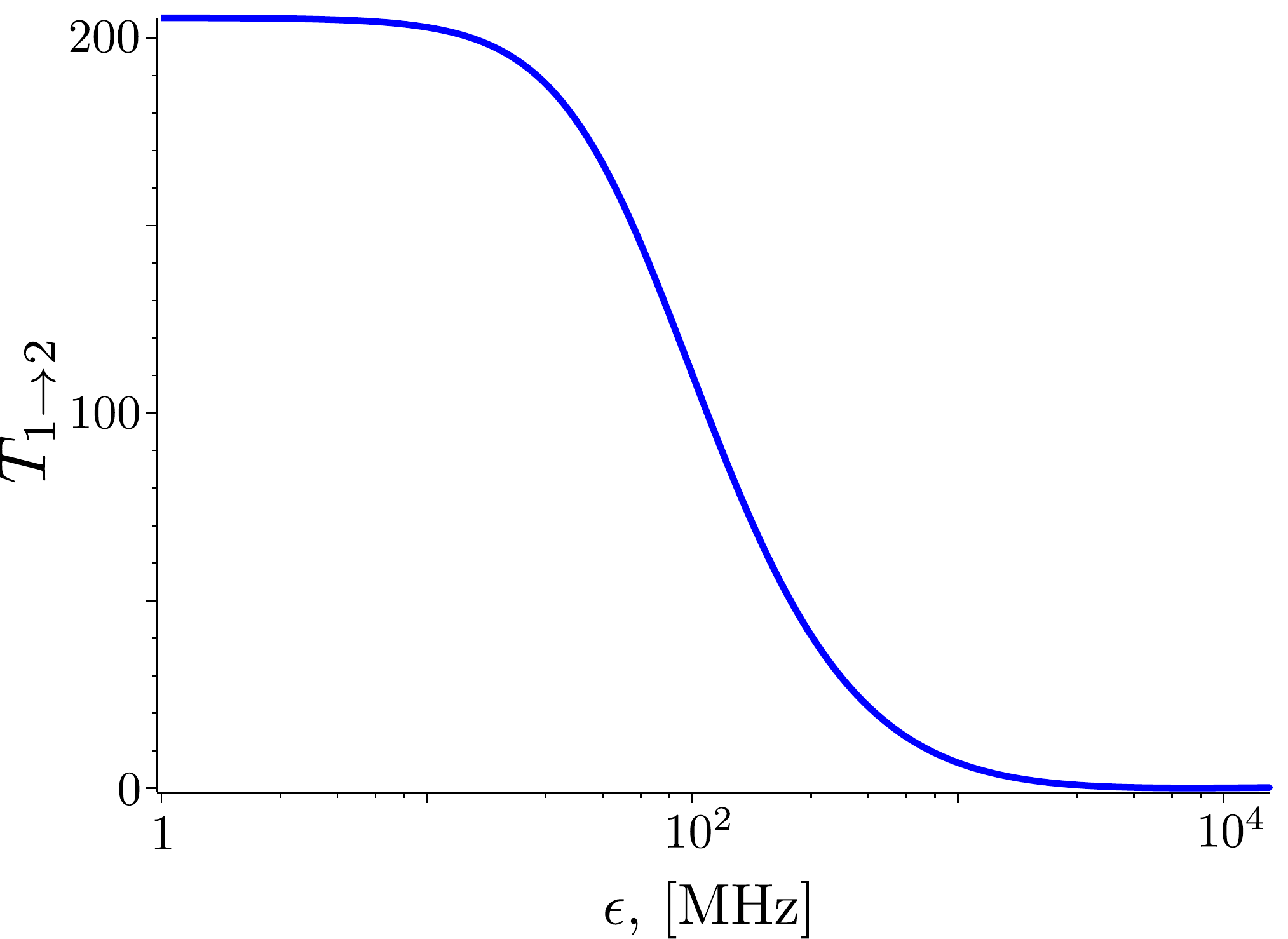}
\caption{{Transmissivity $T_{1\rightarrow2}$ of light as a
function of the input driving field signal $\epsilon$, with
resonant driving frequency $\omega=\omega_c$, when there is no
passive cavity in the system, i.e., $\kappa=0$  (see
Fig.~\ref{fig01}). The driving field coupling constant $\epsilon$
is related to the amplitude $A_{\rm in}$ of the input signal via
the expression $\epsilon=\sqrt{\gamma_1}A_{\rm in}$.  The plot
presents the gain behaviour in the active cavity versus the
intensity of the input signal field. With increasing strength of
the input signal $\epsilon$, the gain (represented by the
transmissivity $T_{1\rightarrow2}$) of the signal field steadily
decreases due to gain saturation. The remaining parameters used
here are the same as in Fig.~\ref{fig03}. For details, regarding
the transmission coefficients, see also Sec.~III~E.}}\label{fig08}
\end{figure}

{We plotted Figs.~\ref{fig05}--\ref{fig07} to demonstrate the
aforementioned asymmetry of the steady-state intensities in both
cavities depending on the propagation direction of the resonant
signal field. As it follows from Fig.~\ref{fig07}, for very small
values of $\kappa<\kappa_{\rm EP}$, the field intensity in the
active cavity for the direction $4\rightarrow1$ can be two or
three orders of magnitude larger than the intensity in the passive
cavity for the opposite propagation direction $1\rightarrow4$. As
a consequence, this asymmetric property can lead to the
observation of non-reciprocal light behavior for small $\kappa$.
Therefore, the non-reciprocity of light propagation, as
experimentally demonstrated in, e.g., Ref.~\cite{Peng2014}, arises
from the combination of loss, gain, and gain saturation.
Specifically, owing to the saturation effects, the gain
experienced by the signal entering the amplifying cavity strongly
depends on its intensity. We would like to stress again that the
input signal with a large intensity undergoes a much lower
amplification with respect to a weaker signal due to gain
saturation. Hence, feeding the gain cavity with a quite strong
field gives rise to a modest amplification (see also
Fig.~\ref{fig08}). This signal is finally strongly absorbed by the
lossy cavity, before being detected. By contrast to this, if the
same signal is first sent to the lossy cavity, then it is strongly
absorbed before entering the gain cavity. Such a small signal does
not saturate the gain medium and is strongly amplified before
detection.}


{As Figs.~\ref{fig03} and \ref{fig04} also indicate, { in unbroken $\cal PT$-symmetry phase,} for large
values of the intercavity coupling $\kappa\gg\kappa_{\rm EP}$,  the
system exhibits a linear character, regardless of both strengths of
the input signal field and the propagation direction
($1\rightarrow4$ and $4\rightarrow1$). Indeed, in this case, the
coupling $\kappa$ between two microcavities becomes large enough,
enabling the input signal fields to freely propagate in either
direction. Hence, there is no localization of the fields in the
system, and, thus, there is no observed nonlinearity due to gain
saturation. }

\subsection{Transmission spectra}

Here we focus on the spectral properties of the driving fields
that propagate through the system.

By rewriting the complex amplitudes {$A_k$ of the fields [as in
\eqref{REC}] as} $A_k=|A_k|\exp({i\phi_k})$, one arrives at a
cubic equation for the field intensity $I_1$ in the active cavity
in the steady state (see Appendix~B, for details):
\begin{equation}\label{CE}  
\lambda_1I_1^3+\lambda_2I_1^2+\lambda_3I_1+\lambda_4=0,
\end{equation}
with coefficients $\lambda_i$ defined as:
\begin{eqnarray}\label{lambda}  
&\lambda_1 = \frac{B^2}{4}, \quad \lambda_2 = BF, \quad \lambda_4 = -\epsilon^2,& \nonumber \\
&\lambda_3 = F^2+\Delta^2\left(f-1\right)^2, \quad F=\frac12\left(f\Gamma_2-G_1\right),& \nonumber \\
\end{eqnarray}
where $f=4\kappa^2/(\Gamma_2^2+4\Delta^2)$. Equation~(\ref{CE})
has only one real solution (see Appendix~B, for details), when its
discriminant is negative, which is always the case when, e.g.,
$A\approx \Gamma_1$ regardless of $\Gamma_2$ and $\kappa$.

The transmission spectrum can be calculated as follows
\begin{equation}\label{T}  
T(\omega)=\left|\frac{A_{\rm out}}{A_{\rm in}}\right|^2.
\end{equation}
\begin{figure}[tb!]  
\includegraphics[width=0.49\textwidth]{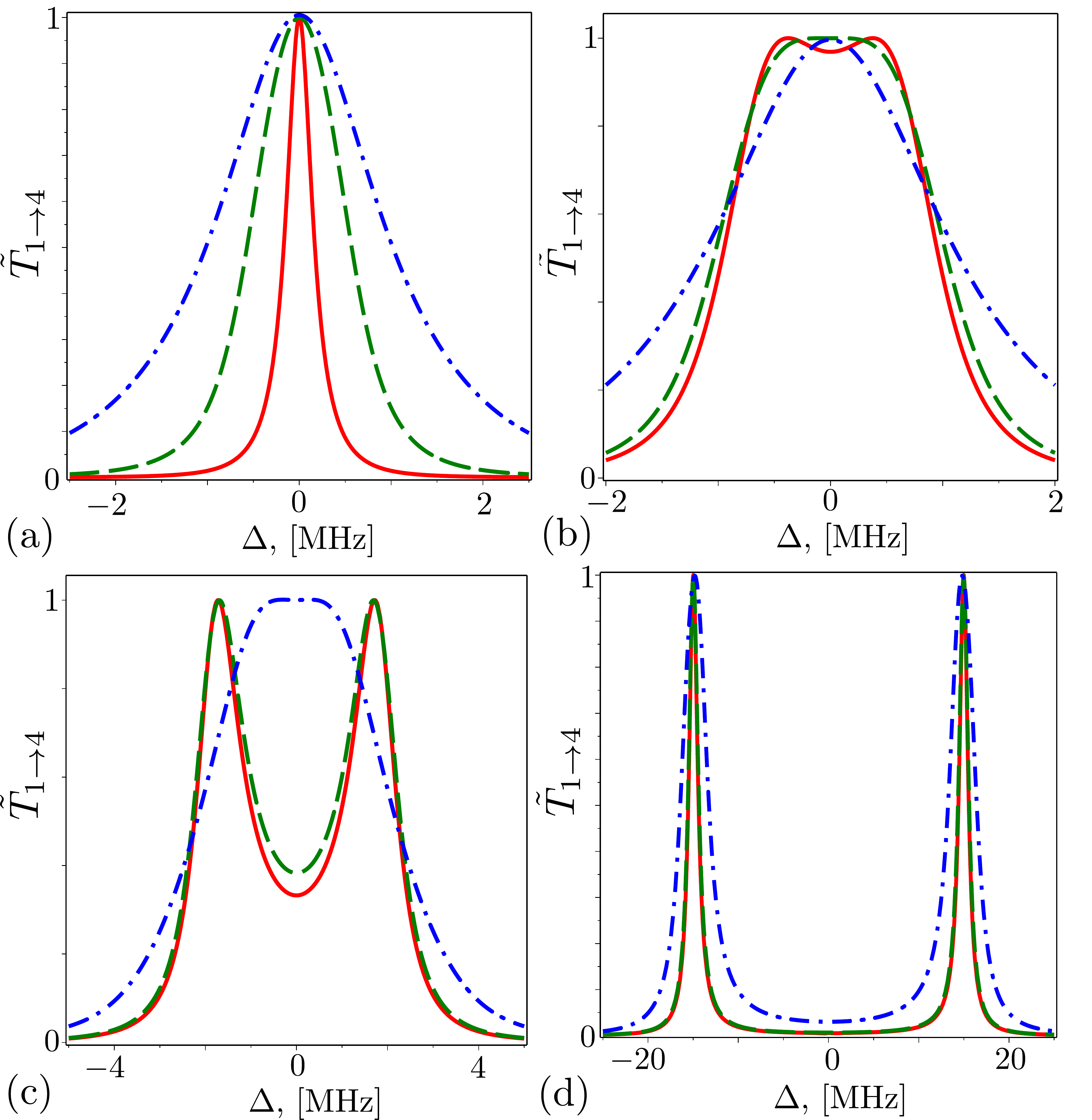}
\caption{Normalized transmission spectrum  $\tilde T_{1\rightarrow
4}= T_{1\rightarrow 4}(\Delta)/ {\rm max}[T_{1\rightarrow
4}(\Delta)]$ versus detuning $\Delta=\omega-\omega_c$, where $
T_{1\rightarrow 4}(\Delta)$ is given in \eqref{T14} (see also
Fig.~\ref{fig01}, for details), for different values of the
intercavity coupling $\kappa$: (a) $\kappa=0.3$~MHz, (b) $\kappa=0.85$~MHz, (c)
$\kappa=1.88$~MHz and (d) $\kappa=15$~MHz  with the \PT-symmetry condition $A-C_1-C_2=0$
(excluding the  gain saturation term $BI_1$). The linear regime
with $B=0.05$~Hz and $\epsilon=1$~MHz (red solid curve); the
nonlinear regime, according to \eqref{NPT}, with $B=0.05$~Hz and
$\epsilon=2$~GHz (green dashed curve) and $\epsilon=20.5$~GHz
(blue dash-dotted curve). Assuming the passive cavity loss
$C_2=1$~MHz, the active cavity gain $A=301$ MHz, and the
waveguides coupling with both cavities is $\gamma=1.15$~MHz.
{The transmission spectra in panel (a) and panel (d) exhibit Lorentzian line shapes,
when the system is well below or above from the EP, respectively~\cite{Yoo2011}. On the contrary, in panel (b) and 
panel (c), the transmission spectra have squared Lorentzian
line shapes, which is a signature of the occurrence of
EPs~\cite{Yoo2011,Sweeney2018,Pick2017}.} }\label{fig09}
\end{figure}
To obtain the transmission spectrum $T_{1\rightarrow4}(\omega)$ at
port 4 when sending the signal from port 1, one needs to know the
expressions for the corresponding input-output fields. The output
field at port 4 is found as $A_{\rm out}=\sqrt{\gamma_2}A_2$. The
input driving field sent from port 1 can be expressed,  as $A_{\rm
in}=\epsilon/\sqrt{\gamma_1}$, where, again, $\epsilon$ is the coupling
strength between input driving field and microresonator  (Fig.~\ref{fig01}). By rewriting the field $A_2$ via $A_1$ and using
\eqref{REC}, one finally  obtains
\begin{equation}\label{T14}  
T_{1\rightarrow4}(\Delta)=\frac{4\kappa^2\gamma_1\gamma_2}{\epsilon^2(\Gamma_2^2+4\Delta^2)}I_1.
\end{equation}
The same analysis can be carried out for the case when the driving
coherent field is sent from port 4, and the signal is detected at
port 1. In that case, one obtains the same cubic equation for the
field intensity $I_1$, as in \eqref{CE}, but with different
$\lambda_k$; $k=1,\dots,4$ (see Appendix~B). The expression for
the transmission spectra $T_{4\rightarrow1}(\omega)$ then attains
the following form
\begin{equation} \label{T41}
T_{4\rightarrow1}(\omega)=\frac{\gamma_1\gamma_2}{\epsilon^2}I_1.
\end{equation}
Similarly, the transmission spectrum $T_{1\rightarrow2}$ can be
found using the input-output relation: $A_{\rm out}=A_{\rm
in}+\sqrt{\gamma_1}A_1$ (see Appendix~B, for details).

\begin{figure} [tb!]
\includegraphics[width=0.3\textwidth]{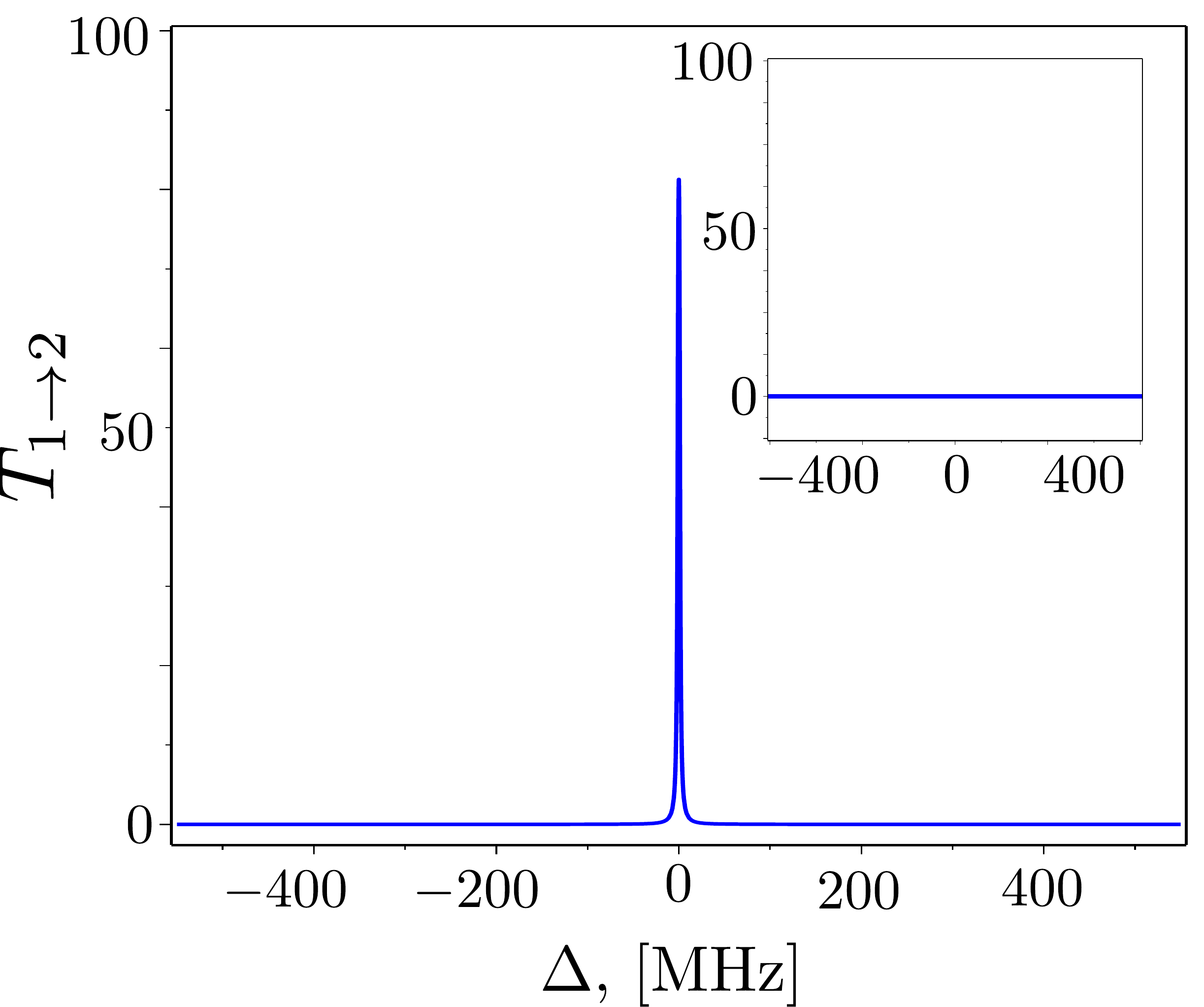}
\caption{Transmission spectrum $T_{1\rightarrow2}$ assuming that there is
 coupling only between the WG$_1$ and the active microcavity $R_1$;
the active cavity gain is $A=20.4$~MHz, and the power of the driving
field is ${P}=100$~nW. The inset shows the transmissivity when the
power of the driving field is ${P}=0$. Here and in the graphs
below, the resonance wavelength of both cavities is set as
$\lambda_c=1550$ nm. Moreover, we assume that all losses in both
cavities are encompassed by the waveguides couplings, i.e.,
$\Gamma_i=C_i+\gamma_i\approx \gamma_i$, $i=1,2$, and their values
are fixed along with the gain saturation coefficient, i.e.,
$\gamma_1=25$~MHz, $\gamma_2=10$~MHz, and $B=0.1$~Hz.  This figure
qualitatively reproduces the experimentally-obtained Fig.~1(f) in
Ref.~\cite{Peng2014}.}\label{fig10}
\end{figure}
\begin{figure}[tb!] 
\includegraphics[width=0.5\textwidth]{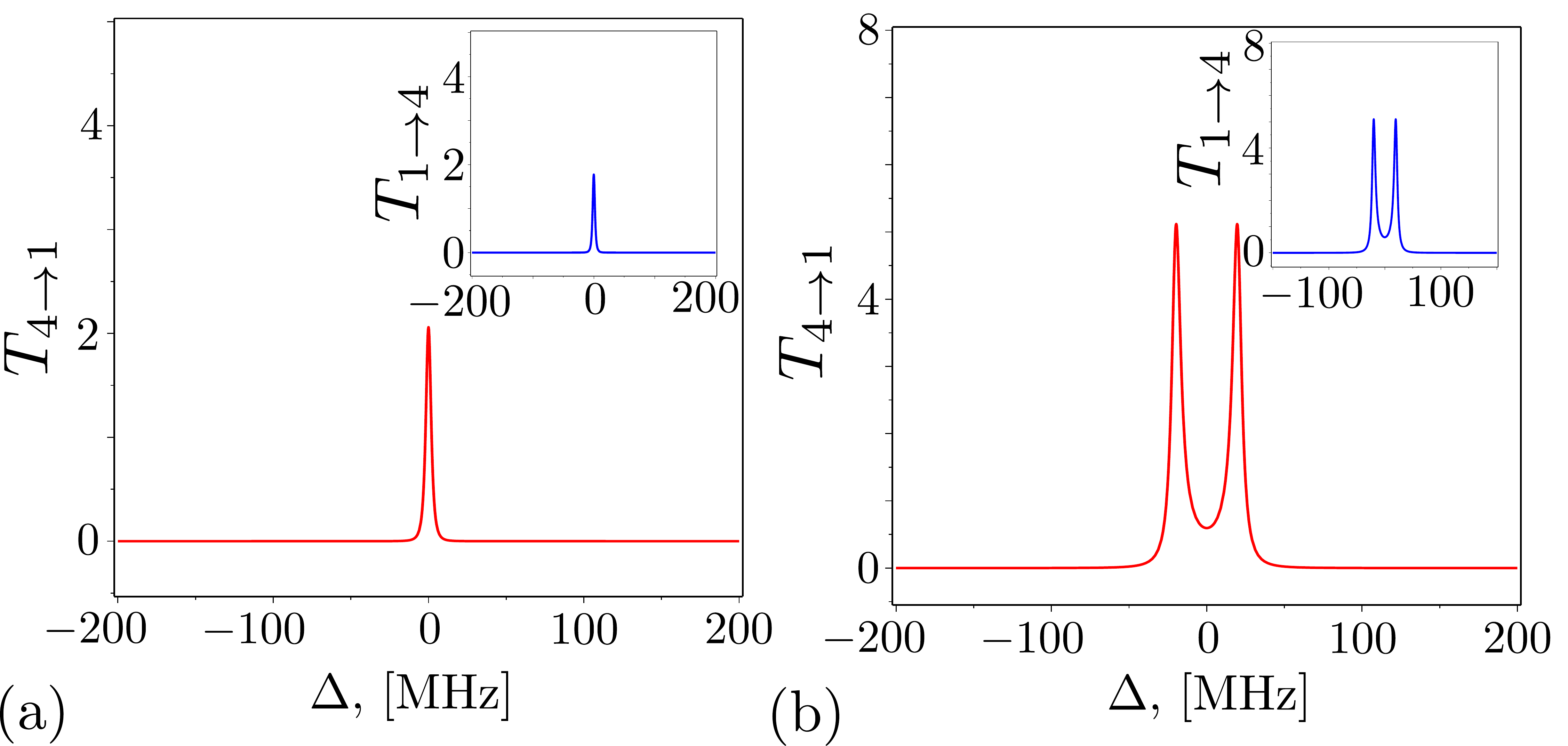}
\caption{Transmission spectrum $T_{4\rightarrow1}$ (red curve) and
$T_{1\rightarrow4}$ (blue curve inside in the captions) in the
linear regime when: (a) $ A=21$~MHz, $\kappa=1$~MHz, and
${P}=100$~nW; (b) $ A=21$~MHz, $\kappa=20$~MHz, and ${P}=100$~nW.
This figure qualitatively reproduces the experimentally-obtained
Fig.~3 in Ref.~\cite{Peng2014}}\label{fig11}
\end{figure}
\begin{figure*}[tb!] 
\includegraphics[width=\textwidth]{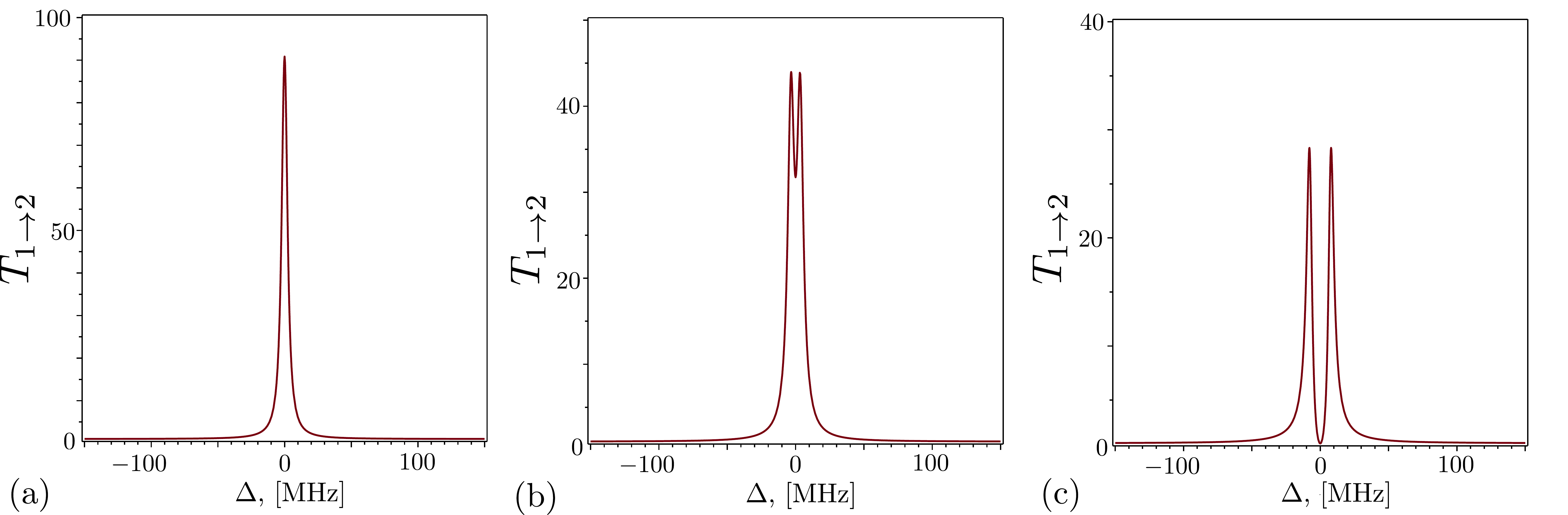}
\caption{Transmission spectrum $T_{1\rightarrow2}$ when there is a
coupling between the microresonators $R_1$ and $R_2$ for: (a)
$A=21$~MHz, $\kappa=1$ MHz; (b) $A=24$~MHz, $\kappa=4$~MHz;
and (c) $A=25.5$~MHz, $\kappa=8$~MHz. The power of the input
signal is ${P}=100$~nW.  This figure qualitatively reproduces the
experimentally-obtained Fig.~S6 in
Ref.~\cite{Peng2014}.}\label{fig12}
\end{figure*}
\begin{figure*} [tb!]
\includegraphics[width=\textwidth]{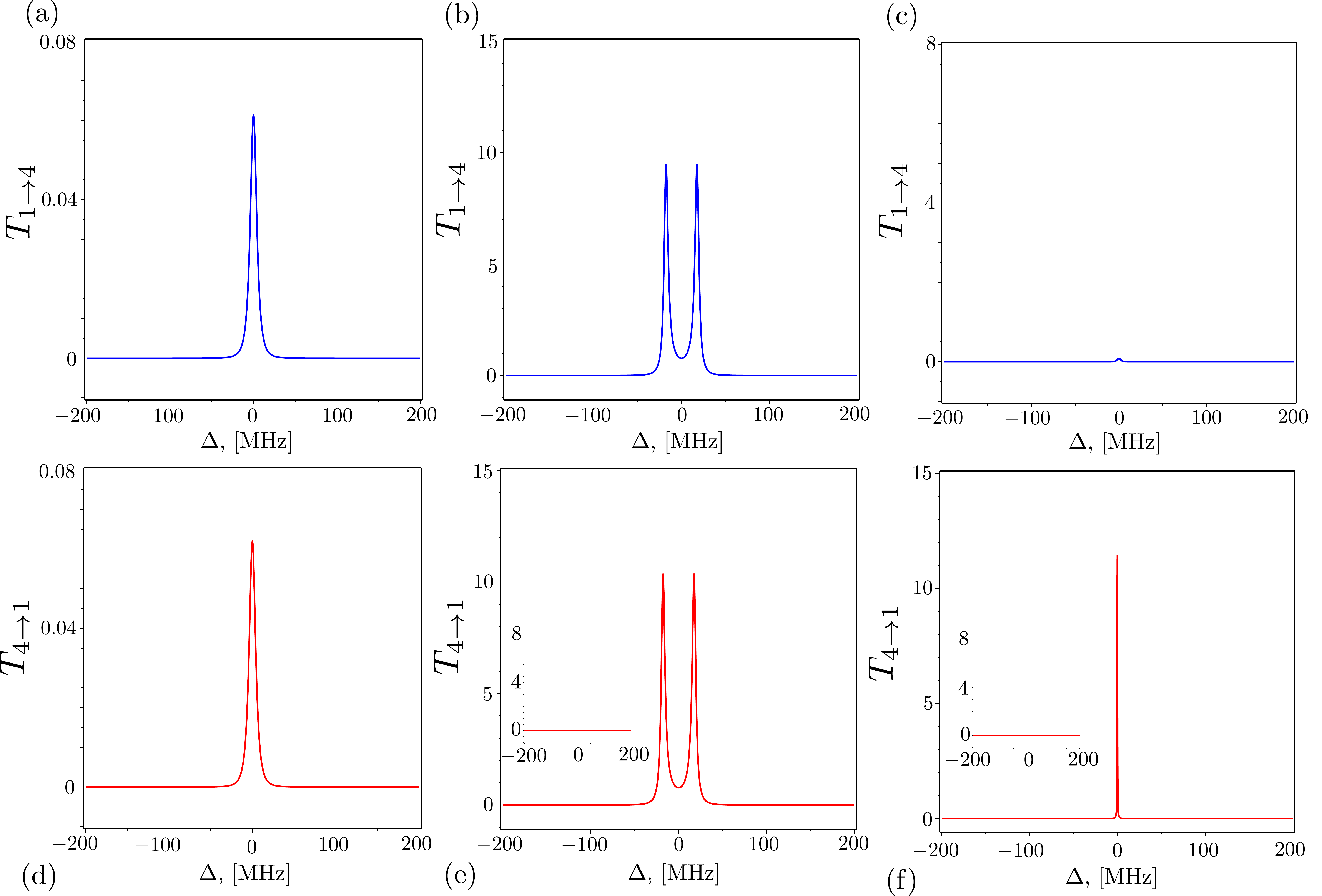}
\caption{Transmission spectrum $T_{1\rightarrow4}$
$\{T_{4\rightarrow1}\}$ versus detuning $\Delta=\omega-\omega_c$
for: (a)  $\{$(d)$\}$ no amplification $A=0$~MHz, $\kappa=1$~MHz;
(b)  $\{$(e)$\}$ $A=25$~MHz, $\kappa=20$~MHz; (c) $\{$(f)$\}$
$A=25$~MHz, $\kappa=0.2$~MHz.  The power of the input signal is
${P}=1$ $\mu$W. {Graphs (c) and (f) clearly show non-reciprocality
of light propagation.} The insets of panels (e) and (f) show the
spectra assuming no input signal. This figure qualitatively
reproduces the experimentally-obtained Fig.~4 in
Ref.~\cite{Peng2014}.}\label{fig13}
\end{figure*}

As an example,  in Fig.~\ref{fig09} we plot the normalized
transmission spectrum $\tilde T_{1\rightarrow 4}= T_{1\rightarrow
4}(\Delta)/ {\rm max}[T_{1\rightarrow 4}(\Delta)]$ for the
\PT-symmetric condition $A-C_1-C_2=0$, for different values of the
intercavity coupling $\kappa$, and for different intensities of
the driving field (i.e., by varying $\epsilon$) in both linear
($BI_1/C_2\ll 1$) and nonlinear ($BI_1/C_2\approx 1$) regimes [for
details regarding the nonlinearity condition, see \eqref{NPT}].
One can observe an increasing spectral line broadening of the
transmitted light  for stronger driving fields $\epsilon$ in the
nonlinear regime, indicating rising losses due to the gain
saturation for small values of $\kappa$ [green dashed and blue
dash-dotted curves  in Fig.~\ref{fig09}(a)]. {Moreover, the
transmission spectra in Figs.~\ref{fig09}(b) and \ref{fig09}(c) display a squared
Lorentzian line shape, which is characteristic for
EPs~\cite{Yoo2011,Sweeney2018,Pick2017}.} Also, in the nonlinear case, the
splitting of the supermodes occurs for larger $\kappa>\kappa_{\rm
EP}$  and more intensive fields [Figs.~\ref{fig09}(b), \ref{fig09}(c)]. A similar
behavior is observed in the transmission spectrum $T_{4\rightarrow
1}$.


\section{{A comparison with experimental results of
Ref.~\cite{Peng2014}}}

For simplicity, when plotting the graphs in this section, we
assume that the losses in both cavities are comprised mainly by
the losses due to the coupling of the cavities to the waveguides
$\Gamma_i=C_i+\gamma_i=\gamma_i$, $i=1,2$, i.e., we set the
intrinsic losses $C_i$ to zero. The latter assumption also implies
that the system considered has a broken \PT-symmetry, i.e.,
$A-C_1-C_2\neq0$, according to \eqref{pt_tot2}. At the same time,
the total losses in the system are expected to be larger than the
gain, i.e., $(A-\gamma_1-\gamma_2)<0$. Also, the active
microcavity is assumed to operate near the threshold $A\approx
\gamma_1>\gamma_2$.

In this section, we discuss  possible applications of the
semiclassical Scully-Lamb laser theory  in the prediction of some
nontrivial light behavior that was experimentally observed in
Refs.~\cite{Peng2014,Chang2014}. In those papers, the authors
experimentally studied a system of coupled \PT-symmetric
whispering-gallery microcavities, i.e., a system that is identical
to that presented in Fig.~1 and which is the focus of the
theoretical study of this work. Below, we theoretically
reproduce some of the experimental  graphs of Ref.~\cite{Peng2014}
{in a qualitative rather than quantitative way (meaning that
we make some additional assumptions regarding the system
parameters, used in constructing the graphs here).} Nevertheless,
as was just mentioned, a qualitative comparison can be made, and
positive conclusions can be inferred regarding the applicability
of the {Scully-Lamb laser model, in its semiclassical limit,} to
explain some of the results of Ref.~\cite{Peng2014}. We note that
in the construction of the graphs shown here, which are presented
in Figs.~\ref{fig10}--\ref{fig13}, we do \emph{not} invoke \PT-symmetry
in the system (see also the text below).

For clarity, {in the captions of some of our figures we
indicate the corresponding experimental plots of
Ref.~\cite{Peng2014}} that we try to theoretically reproduce.
Also, in order to stress the similarity between the figures
reproduced here and the original experimental graphs of
Ref.~\cite{Peng2014}, we keep the axis scales of the plots  to
be the same as those given in Ref.~\cite{Peng2014}.

For example, in Fig.~\ref{fig10}, {which corresponds to the
experimentally-obtained Fig.~1(f) in Ref.~\cite{Peng2014},} the
transmission spectrum $T_{1\rightarrow2}$ is shown when  only the
WG1 and the microresonator $R_1$ are coupled in the system and
$\kappa=0$, i.e., only the first resonator is considered. Thus, by
coupling the signal sent from port 1 to the active cavity, one
obtains a substantial signal amplification, when detecting the
output signal at port 2. In this case, the active cavity just
enhances the incoming field. The inset of Fig.~\ref{fig10}
demonstrates that  amplification does not occur  if there is no
input signal.


In Fig.~\ref{fig11}, {which corresponds to the
experimentally-obtained} Fig.~3 in Ref.~\cite{Peng2014}, we show
the transmission spectra for the light propagating in the
directions $1\rightarrow4$ and $4\rightarrow1$ in the linear
regime, i.e., the laser cavity is assumed to be below the lasing
threshold $A<\gamma_1$. One can see the linear response of the
propagating signal when the inter-cavity coupling coefficient
$\kappa$ is lower or larger than $\kappa_{\rm cr}$, where
$\kappa_{\rm cr}$ denotes the critical point, when the supermodes
start splitting. We note that, in Fig.~\ref{fig11}, a linear
behavior of the transmitted light is already observed in the
unbalanced gain-loss regime, i.e., when \PT-symmetry is broken in
the system.

Figure~\ref{fig12}, {which corresponds to the
experimentally-obtained }Fig.~S6 in Ref.~\cite{Peng2014}, displays
the transmitted spectrum $T_{1\rightarrow2}$ versus the detuning
$\Delta=\omega-\omega_c$, for various values of the gain $A$ and
the intercavity coupling $\kappa$. One can see the appearance of
the supermodes splitting in the system with the increasing values
of $\kappa$. Conversely, for smaller values of $\kappa$, the two
supermodes coalesce resulting in only one peak in the spectrum.

\subsection*{Non-reciprocity of light propagation}

{Here, to complement Sec. III~D, we further discuss the
theoretical microscopic prediction of non-reciprocity of the
propagating light in the considered coupled active-passive
microresonators system as shown Fig.~\ref{fig13}, which
corresponds to the experimentally-obtained Fig.~4 in
Ref.~\cite{Peng2014}. It is seen that} there is an enhancement in
the transmitted light from port 4 to port 1, and tending to zero
transmission $T_{1\rightarrow4}$ in the opposite direction for
small values of $\kappa$. Meaning that the system starts behaving
nonreciprocally. The latter nonlinear effect was observed in
Ref.~\cite{Peng2014}, but, naturally, could not be explained based
on the linear rate equations used there. \emph{Utilizing the
semiclassical  laser theory, one can attain the needed nonlinear
term arising from the laser gain saturation} in the active
microcavity. Moreover, \emph{this non-reciprocity is observed
without invoking \PT-symmetry}, because it can be already observed
when the gain and loss are unbalanced in the system (see
Fig.~\ref{fig13}). {The inset of Fig.~\ref{fig13}(f) indicates
that the observed nonlinearity is not caused by the lasing
initiated by spontaneous emission in the active cavity.}

\section{Conclusions}

We have applied the quantum Scully-Lamb laser theory to a pair of
\PT-symmetric coupled whispering-gallery microcavities, i.e., a
system, which consists of both active and  passive microring
cavities, such that gain and losses are balanced in the system. It
has been shown that, in the nonlinear regime, or more precisely,
under the condition in which the gain saturation in the active
cavity is comparable to the losses in the passive cavity, the
intense intracavity fields of the steady state lead to the
modification of the eigenmodes and of the EPs of the \PT-symmetric
system. Namely, the imaginary part of the eigenspectrum acquires
an extra negative term due to the gain saturation effects. This
effect leads to the shift of the EP either to lower or larger
values depending on the gain saturation $B$ {and the propagation
direction of the driving fields.} Starting from the master
equation for this coupled system, including dissipation, gain, and
gain saturation, and applying the semiclassical approximation, we
are able to describe the experimental results obtained in
Refs.~\cite{Peng2014,Chang2014}. In particular, this approach is
able to reproduce the observed non-reciprocal light propagation in
the coupled system of whispering-gallery microcavities. We have
also shown that the gain saturation mechanism in the active cavity
is crucial for the observation of light non-reciprocity. Moreover,
we have found that the unidirectional light propagation can be
observed even when the \PT-symmetry condition is not fulfilled. It
should be stressed that neither \PT-symmetry nor its breaking is
required for nonreciprocity. The nonreciprocity observed in our
system is a result of a nonlinearity and, in the broken
\PT-regime, such nonlinearity can be observed at much lower input
intensity.

In summary, we proposed, applied, and validated the Scully-Lamb
laser model, with the non-Lindbladian master equation, for coupled
resonators with losses, gain, and gain saturation. Although we
studied non-reciprocity and exceptional points applying the
semiclassical approximation, this master equation allows for a
quantum description of the cavity fields. This approach
constitutes a promising tool for the study of quantum optical
effects in coupled resonators with balanced (or unbalanced) gain
and losses.

\acknowledgements
I.I.A. was supported  by  Grant Agency of the Czech Republic (Project No.~17-23005Y), the Project
CZ.02.1.01\/0.0\/0.0\/16\_019\/0000754, and the Project  LO1305 of the Ministry of Education, Youth
and Sports of the Czech Republic.
A.M. and F.N. acknowledge the support of a grant from the John
Templeton Foundation. F.N. is supported in part by the: MURI
Center for Dynamic Magneto-Optics via the Air Force Office of
Scientific Research (AFOSR) (FA9550-14-1-0040), Army Research
Office (ARO) (Grant No.~W911NF-18-1-0358), Asian Office of Aerospace
Research and Development (AOARD) (Grant No.~FA2386-18-1-4045),
Japan Science and Technology Agency (JST) (the Q-LEAP program, and
CREST Grant No.~JPMJCR1676), Japan Society for the Promotion of
Science (JSPS) (JSPS-RFBR Grant No.~17-52-50023), and the
RIKEN-AIST Challenge Research Fund.
S.S. is supported in part by the Army Research Office (ARO) (Grant No.~W911NF1910065).
S.K.O. is supported by Army Research Office (ARO) grant No.~W911NF-18-1-0043, Air Force
Office of Scientific Research (AFOSR) award No.~FA9550-18-1-0235, National Science
Foundation (NSF) (1807485), and by The Pennsylvania State University, Materials Research
Institute (MRI).


\appendix
\section{Derivation of the master equation in \eqref{MES}}

To make this article self-consistent, in this Appendix, we present
a derivation of the quantum laser master equation, given in
Eq.~(\ref{MES}), {for the Scully-Lamb laser model based on the
derivation of Yamamoto and Imamo\v{g}lu in
Ref.~\cite{YamamotoBook}.} This derivation bears a
phenomenological character and, as such, naturally allows to
incorporate all the terms of the interaction Hamiltonian of the
fields without the need to solve the Schr\"odinger equation
directly. {Another derivation of the master equation~(\ref{MES}) can be found in Ref.~\cite{OrszagBook}.}

\begin{figure}[tb!] 
\includegraphics[width=0.25\textwidth]{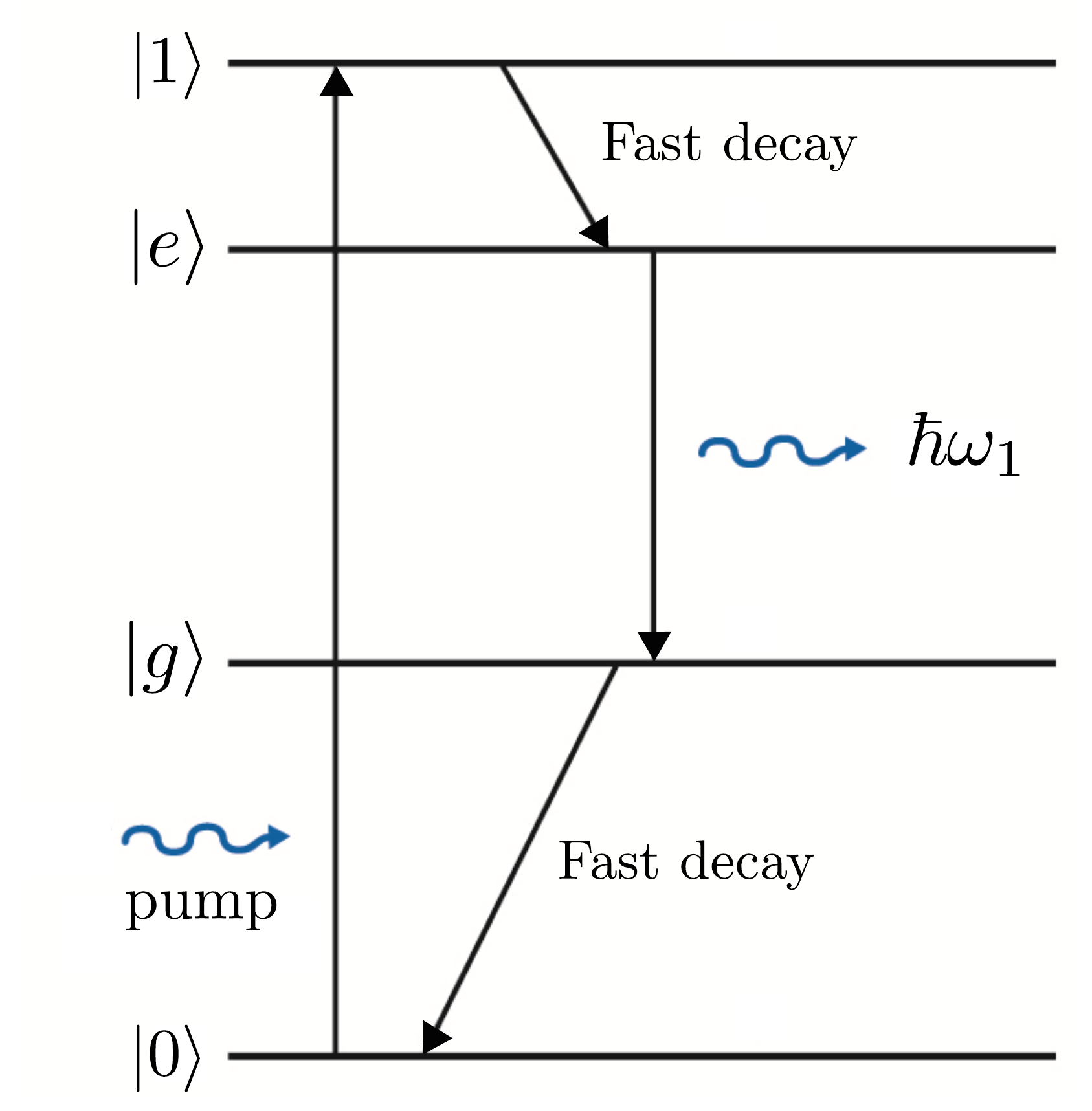}
\caption{Four-level atom laser scheme in an active
microresonator.}\label{fig14}
\end{figure}

The active cavity is represented by a general four-level laser
system in which the two intermediate energy levels are coupled by
the laser mode (see Fig.~\ref{fig14}). In this limit, the uppermost level
of the atom may be adiabatically eliminated to give an effective
three-level system. The latter
assumption is valid as long as the decay rate from the uppermost
state $|1\rangle$ to the upper laser level $|e\rangle$ is much
faster than all other rates in the atom-field system. In this
limit, one has an effective incoherent pumping rate $r$ from the
atomic ground state $|0\rangle$  into $|e\rangle$. Additionally, the laser mode in the
active cavity is coupled to the passive microresonator.

The interaction Hamiltonian $\hat H_I$, describing such two
coupled active-passive microresonators system, is given by
\begin{eqnarray}\label{H1}
\hat H_I = ig\left(\hat\sigma_{+}\hat a_1-\hat\sigma_{-}\hat
a_1^{\dagger}\right)+i\kappa\left(\hat a_1\hat a_2^{\dagger}-\hat
a_1^{\dagger}\hat a_2\right).
\end{eqnarray}
{Specifically, the second term in Eq.~(\ref{H1}) describes the
linear interaction between the modes in the active ($a_1$) and
passive ($a_2$) resonators; while the first term describes the
interaction between the mode $a_1$ with only two levels
($|g\rangle$ and $|e\rangle$)
within the standard Jaynes-Cummings
model.} The operators $\hat\sigma_k$, $k=+,-$, are spin-raising
and spin-lowering operators of the atom in the active medium,
respectively. The constants $g$, $\kappa$ denote the coupling
strength between the atom and the field in the laser cavity, and
between the two fields propagating in two different cavities,
respectively.  We also assumed that the atomic and active cavity
field resonances coincide.

The quantum Liouville equation for the density operator $\hat\rho$
in the interaction picture, which describes the atom-field-field
dynamics is:
\begin{equation}\label{ME1}
\frac{d}{dt}\hat\rho=\frac{1}{i\hbar}\left[\hat
H_{I},\hat\rho\right].
\end{equation}
In the active cavity, the optical gain is provided by the excited
atoms, which are pumped by an external field. The interaction
between the atom and an external pumping field, that provides an
inverse population in our effective three-level atom laser
(depicted in Fig.~\ref{fig14}), can be described by the
following equation
\begin{equation}\label{ME2}
\frac{d}{dt}\hat\rho=-\frac{r}{2}\left(\hat\sigma_{00}\hat\rho+\hat\rho\hat\sigma_{00}-2\hat\sigma_{e0}\hat\rho\hat\sigma_{0e}\right),
\end{equation}
where $\hat\sigma_{00}=\hat\sigma_{0e}\hat\sigma_{e0}$, and
$\hat\sigma_{0e}$ ($\hat\sigma_{e0}$) is the spin operator for the
{atomic transition from $|0\rangle$ ($|e\rangle$) to
$|e\rangle$ ($|0\rangle$).} The coefficient $r$ accounts for the
pumping rate of the atom.

To include spontaneous emission and the  emission caused by
external dephasing processes into an external reservoir field, we
need to add in \eqref{ME1} the following terms:
\begin{eqnarray}\label{ME3} 
\frac{d}{dt}\hat\rho &=& -\frac{\gamma_{\rm sp}}{2}\left(\hat\sigma_{+}\hat\sigma_{-}\hat\rho+\hat\rho\hat\sigma_{+}\hat\sigma_{-}-2\hat\sigma_{-}\hat\rho\hat\sigma_{+}\right) \nonumber \\
&&-\frac{\gamma_d}{2}\left(\hat\sigma_{ee}\hat\rho\hat\sigma_{gg}+\hat\sigma_{gg}\hat\rho\hat\sigma_{ee}\right),
\end{eqnarray}
where $\gamma_{\rm sp}$ and $\gamma_d$, are, respectively, the
rates of spontaneous emission, and the emission imposed by additional
dephasing processes.

Now, collecting together Eqs.~(\ref{ME1})--(\ref{ME3}), one
arrives at the master equation for the density operator $\hat\rho$
\begin{eqnarray}\label{MEG} 
\frac{d}{dt}\hat\rho &=& \frac{1}{i\hbar}\left[\hat H_{I},\hat\rho\right] - \frac{1}{2}\sum\limits_{i=1}^{2}C_i\hat L^d_i(\hat\rho) \nonumber \\
&&-\frac{r}{2}\left(\hat\sigma_{00}\hat\rho+\hat\rho\sigma_{00}-2\hat\sigma_{e0}\hat\rho\hat\sigma_{0e}\right) \nonumber \\
&&-\frac{\gamma_{\rm sp}}{2}\left(\hat\sigma_{+}\hat\sigma_{-}\hat\rho+\hat\rho\hat\sigma_{+}\hat\sigma_{-}-2\hat\sigma_{-}\hat\rho\hat\sigma_{+}\right)\nonumber \\
&&-\frac{\gamma_d}{2}\left(\hat\sigma_{ee}\hat\rho\hat\sigma_{gg}+\hat\sigma_{gg}\hat\rho\hat\sigma_{ee}\right),
\end{eqnarray}
where we introduced a Lindbladian damping super operator as $\hat
L^d_i(\hat O)=\hat a_i^{\dagger}\hat a_i\hat O+\hat O\hat
a_i^{\dagger}\hat a_i-2\hat a_{i}\hat O\hat a_i^{\dagger}$.

To obtain the master equation for the reduced field density
operator $\hat\rho_f$, which describes the dynamics of the optical
fields in the cavities, one has to trace out the density operator
$\hat\rho$ in \eqref{MEG} over the atom states. Namely,
\begin{equation} 
\frac{d}{dt}\hat\rho_f={\rm
Tr}_a\left[\frac{d}{dt}\hat\rho\right]=\sum\limits_{i=0,e,g}\left\langle
i\Big|\frac{d}{dt}\hat\rho\Big|i\right\rangle.
\end{equation}

\begin{eqnarray}\label{rhof1}  
\frac{d}{dt}\hat\rho_f&=&\frac{1}{i\hbar}\left[\hat H_{f},\hat\rho_f\right] - \frac{1}{2}\sum\limits_{i=1}^{2}C_i\hat L^d_i(\hat\rho_f)\nonumber \\
&&+g(\hat a_1\hat\rho_{ge}-\hat a_1^{\dagger}\hat\rho_{eg}-\hat\rho_{ge}\hat
a_1+\hat\rho_{eg}\hat a_1^{\dagger}),
\end{eqnarray}
where the Hamiltonian $H_f$ accounts for the field interaction
between active and passive cavities, i.e., it is the second term
in Eq.~(\ref{H1}). The operator $\hat\rho_{ge}=\left\langle
g\Big|\frac{d}{dt}\hat\rho\Big|e\right\rangle$, from \eqref{MEG},
obeys the following equation
\begin{equation}\label{ge} 
\frac{d}{dt}\hat\rho_{ge}=g(\hat\rho_{gg}\hat a_1^{\dagger}-\hat
a_1^{\dagger}\hat\rho_{ee})-\frac{\gamma_T}{2}\hat\rho_{ge},
\end{equation}
where $\gamma_T= \gamma_{\rm sp}+\gamma_d$ is the total decay rate
of the atom. The same relation, given in \eqref{ge}, also holds
true for the operator $\hat\rho_{eg}=\langle e|\hat\rho|g\rangle$.

Assuming $\gamma_T\gg C_1, C_2, \kappa$, one eliminates
$\hat\rho_{ge}$ and $\hat\rho_{eg}$  in the adiabatic
approximation, i.e.,  $\frac{d}{dt}\hat\rho_{ge}=0$. Thus, one
 obtains
\begin{equation}\label{rhoge1} 
\hat\rho_{ge}=\frac{2g}{\gamma_T}(\hat\rho_{gg}\hat
a_1^{\dagger}-\hat a_1^{\dagger}\hat\rho_{ee})
\end{equation}

Substituting $\hat\rho_{ge}$ and $\hat\rho_{eg}$ in \eqref{rhoge1}
into \eqref{rhof1} one obtains
\begin{eqnarray}\label{rhof2} 
\frac{d}{dt}\hat\rho_f&=&\frac{1}{i\hbar}\left[H_{f},\hat\rho_f\right]- \frac{1}{2}\sum\limits_{i=1}^{2}C_i\hat L^d_i(\hat\rho_f)\nonumber \\
&&-\frac{2g^2}{\gamma_T}\left(\hat L_1^d(\hat\rho_{gg})+\hat
L_1^a(\hat\rho_{ee})\right),
\end{eqnarray}
where $L_i^a(\hat O)=\hat a_i\hat a_i^{\dagger}\hat O+\hat O\hat
a_i\hat a_i^{\dagger}-2\hat a_{i}^{\dagger}\hat O\hat a_i$ is the
Lindbladian amplification super operator.

Moreover,  $\hat\rho_{ee}$ and $\hat\rho_{gg}$ satisfy the
following equations
\begin{eqnarray}\label{rhoge2}  
\frac{d}{dt}\hat\rho_{ee}&=&\frac{2g^2}{\gamma_T}\left(2\hat a_1\hat\rho_{gg}\hat a_1^{\dagger}-\hat a_1\hat a_1^{\dagger}\hat\rho_{ee}-\hat\rho_{ee}\hat a_1\hat a_1^{\dagger}\right)-\gamma_{sp}\hat\rho_{ee}\nonumber \\
&&+r\hat\rho_{00}, \nonumber \\
\frac{d}{dt}\hat\rho_{gg}&=&\frac{2g^2}{\gamma_T}\left(2\hat a_1^{\dagger}\hat\rho_{ee}\hat a_1-\hat a_1^{\dagger}\hat a_1\hat\rho_{gg}-\hat\rho_{gg}\hat a_1^{\dagger}\hat a_1\right)
+\gamma_{sp}\hat\rho_{ee}. \nonumber \\
\end{eqnarray}
If  the population of the lower energy level $|g\rangle$ is very
low, i.e., the energy quickly decays into the ground state
$|0\rangle$, and if the  gain saturation is weak, then one may
write
\begin{equation}\label{rggr00}  
\hat\rho_{gg}\simeq0, \quad \text{and} \quad
\hat\rho_{00}\simeq\hat\rho_f,
\end{equation}
Applying standard perturbation techniques to \eqref{rhoge2}, one
arrives at
\begin{equation}\label{rhoee} 
\hat\rho_{ee}\simeq\frac{r}{\gamma_{sp}}\hat\rho_f-\frac{2g^2r}{\gamma_T\gamma_{sp}^2}\left(\hat
a_1\hat a_1^{\dagger}\hat\rho_f+\hat\rho_f\hat a_1\hat
a_1^{\dagger}\right).
\end{equation}

Combining now Eqs.~(\ref{rggr00}), (\ref{rhoee}), and
\eqref{rhof2}, one attains the master equation for the two field
operator in the interaction picture
\begin{eqnarray}\label{MEF}  
\frac{d}{dt}\hat\rho_f&=&\frac{1}{i\hbar}\left[\hat H_{f},\hat\rho_f\right]-\frac{A}{2}\hat L^a_1(\hat\rho_f) -\frac{1}{2}\sum\limits_{i=1}^{2}C_i\hat L^d_i(\hat\rho_f) \nonumber \\
&&+\frac{B}{2}\Big[(\am\ap)^2\hat\rho_f+2\am\ap\hat\rho_f\am\ap+\hat\rho_f(\am\ap)^2 \nonumber \\
&&-2\ap\hat\rho_f\am\ap\am-2\ap\am\ap\hat\rho_f\am\Big],
\end{eqnarray}
 with
\begin{equation}
A=\frac{4g^2r}{\gamma_T\gamma_{sp}}, \quad B=\frac{A^2}{2r},
\end{equation}
where the coefficient $A$ stands for the linear gain, and $B$ is
the gain saturation  coefficient.

For the case of the ideal laser system with
$\gamma_{d}=\gamma_{sp}=0$,
one can attain the quantum laser master equation within the
Scully-Lamb laser theory. In the weakly saturated regime, the
spontaneous emission on the laser transition can be discarded, and
one  obtains~\cite{YamamotoBook}
\begin{eqnarray}\label{SL}  
\hat\rho_{gg}&\simeq&\frac{2g^2}{\Gamma^2}\hat a_1^{\dagger}\hat\rho_{ee}\hat a_1=\frac{2rg^2}{\Gamma^3}\hat a_1^{\dagger}\hat\rho_{f}\hat a_1, \nonumber \\
\hat\rho_{ee}&\simeq&\frac{r}{\Gamma}\hat\rho_f-\frac{rg^2}{\Gamma^3}\left[\hat
a_1\hat a_1^{\dagger}\hat\rho_f+\hat\rho_f\hat a_1\hat
a_1^{\dagger}\right],
\end{eqnarray}
where $\Gamma$ is a total decay rate of the states $|g\rangle$ and $|e\rangle$  into the ground state $|0\rangle$ of the atom. Substituting equations in Eqs.~(\ref{SL}) into \eqref{rhof2}, we obtain
\begin{eqnarray}\label{MESL}   
&&\frac{d}{dt}\hat\rho_f=\frac{1}{i\hbar}\left[\hat H_f,\hat\rho_f\right]-\frac{A}{2}\hat L^a_1(\hat\rho_f) -\frac{1}{2}\sum\limits_{i=1}^{2}C_i\hat L^d_i(\hat\rho_f) \nonumber \\
&&+\left[\frac{B}{8}\left\{\hat\rho_f(\am\ap)^{2}+3\am\ap\hat\rho_f\am\ap-4\ap\hat\rho_f\am\ap\am\right\} + \rm {h. c.,}\right], \nonumber \\
\end{eqnarray}
where the gain and  gain saturation coefficients are now expressed
as
\begin{equation}\label{ABcoef}
A=\frac{2g^2r}{\Gamma^2}, \quad \text{and} \quad
B=\frac{4g^2}{\Gamma^2}A.
\end{equation}
{Note a typo in the prefactor of the gain saturation coefficient
$B$ in Eq.~(\ref{ABcoef}) in Ref.~\cite{YamamotoBook}. Namely,
there is a prefactor~1, instead of~4.}

We may also add the Hamiltonian term related to the coupling
between the cavity field $\am$ and the external driving coherent
classical field into the master equation. Such a Hamiltonian can
be given by
\begin{equation}
 \hat H_{\rm drv}=i\epsilon\left[\hat a_1\exp(i\omega_lt)-\hat
a_1^{\dagger}\exp(-i\omega_lt)\right],
\end{equation}
where the coupling constant
$\epsilon\equiv\sqrt{\gamma_1{P}/(\hbar\omega_l)}$ accounts for
the coupling between the driving external coherent field with
power $\cal P$ and the cavity field $\hat a_1$.

By rewriting the  field Hamiltonian $\hat H_f$ in the
Schr\"odinger picture, we finally arrive at \eqref{MES}.

\section{Derivation of \eqref{CE}, and some exact formulas from Sec.~III~E}

Working in the reference frame where the phase of the driving
field is zero,  and by expressing the complex amplitudes in the
rate equations in Eq.~(\ref{REC}) as $A_k=|A_k|e^{i\phi_k}$, one
obtains the following equations for the steady-state:
\begin{eqnarray}\label{B1}  
&i\Delta |A_1|+\frac{G_1}{2}|A_1|
-\kappa |A_2|e^{i(\phi_2-\phi_1)} -\frac{B}{2}|A_1|^3
-\epsilon e^{-i\phi_1}=0,& \nonumber \\
&i\Delta |A_2|-\frac{\Gamma_2}{2}|A_2|+\kappa
|A_1|e^{-i(\phi_2-\phi_1)}=0.&
\end{eqnarray}
Replacing now $|A_2|\exp[i(\phi_2-\phi_1)]$ by $|A_1|$ in the
second equation of \eqref{B1} and inserting it into the first
equation, we attain
\begin{equation}\label{B2}  
\left(i\Delta+\frac{G_1}{2}-\frac{2\kappa^{2}}{\Gamma_2-2i\Delta}\right)|A_1|
-\frac{B}{2}|A_1|^3-\epsilon \exp\left(-i\phi_1\right)=0.
\end{equation}
Separating the real and imaginary parts in \eqref{B2} and
equalizing them to zero, one arrives at
\begin{eqnarray}\label{B3}  
\cos\phi_1&=&\frac{|A_1|}{2\epsilon}\left(G_1-f\Gamma_2-B|A_1|^2\right), \nonumber \\
\sin\phi_1&=&\frac{|A_1|\Delta}{\epsilon}\left(f-1\right).
\end{eqnarray}
where $f$ is given in Eq.~(\ref{lambda}). {Utilizing now the
standard trigonometric relation $\cos^2\phi_1+\sin^2\phi_1=1$,}
and collecting together the coefficients at each order of the real
amplitude, we finally obtain the cubic equation for the field
intensity $I_1=|A_1|^2$, given in \eqref{CE}.

{The unique  real solution of the steady-state intensity
$I_1$, given in  Eq.~(\ref{CE}), is
\begin{equation}\label{B4}  
I_1=\frac{1}{6\lambda_1x}\left[x^2-2\lambda_2x+4\lambda_2^2-12\lambda_1\lambda_3\right],
\end{equation}
where
\begin{eqnarray}  
x^3 = &&12\sqrt{3}\Big(27\lambda_1^2\lambda_4^2-18\lambda_1\lambda_2\lambda_3\lambda_4+4\lambda_1\lambda_3^3+4\lambda_1^3\lambda_4 \nonumber \\
&&-\lambda_2^2\lambda_3^2\Big)^{1/2}+36\lambda_1\lambda_2\lambda_3-108\lambda_1^2\lambda_4-8\lambda_2^3,
\end{eqnarray}
with the coefficients $\lambda_k$ introduced in
Eq.~(\ref{lambda}). For the transmission spectra
$T_{4\rightarrow1}$,  given in Eq.~(\ref{T41}), the steady-state
intensity $I_1$ has the same solution as in Eq.~(\ref{B4}), and
with the same coefficients $\lambda_k$ in Eq.~(\ref{lambda}),
except the coefficient $\lambda_4$, which reads now as
\begin{equation} 
\lambda_4=-f\epsilon^2.
\end{equation}
By combining together the input-output relation $A_{\rm
out}=A_{\rm in}+\sqrt{\gamma_1}A_1$ with the solution for the
phase of the complex amplitude $A_1$ in Eq.~(\ref{B3}), one can
straightforwardly find the transmission spectrum
$T_{1\rightarrow2}$, after applying Eq.~(\ref{T}), as follows
\begin{equation} 
T_{1\rightarrow2}=1+\frac{2\gamma_1I_1}{\epsilon^2}\left(\frac{\gamma_1}{2}-F\right)-\frac{\gamma_1BI_1^2}{\epsilon^2},
\end{equation}
where  $F$ is given in Eq.~(\ref{lambda}). }

%

\end{document}